%% file: ArXivResub.tex
\documentclass[10pt,letterpaper]{article}
\usepackage[top=0.85in,left=2.75in,footskip=0.75in]{geometry}

\usepackage{amsmath,amssymb}

\usepackage{changepage}

\usepackage[utf8x]{inputenc}

\usepackage{textcomp,marvosym}

\usepackage{cite}

\usepackage{nameref,hyperref}

\usepackage[right]{lineno}

\usepackage{microtype}
\DisableLigatures[f]{encoding = *, family = * }

\usepackage[table]{xcolor}

\usepackage{array}

\newcolumntype{+}{!{\vrule width 2pt}}

\newlength\savedwidth

\usepackage{graphicx}

\usepackage{epstopdf}
\usepackage{cite}
\usepackage{color}
\usepackage{placeins}

\usepackage{lscape}

\usepackage[labelfont=bf,labelsep=period,justification=raggedright]{caption}

\makeatletter
\renewcommand{\@biblabel}[1]{\quad#1.}
\DeclareMathOperator*{\argmin}{arg\,min}
\makeatother

\date{}

\raggedright
\usepackage[aboveskip=1pt,labelfont=bf,labelsep=period,justification=raggedright,singlelinecheck=off]{caption}

\makeatletter
\renewcommand{\@biblabel}[1]{\quad#1.}
\makeatother

\date{}

\usepackage{lastpage,fancyhdr,graphicx}
\usepackage{epstopdf}
\fancyheadoffset[L]{2.25in}
\fancyfootoffset[L]{2.25in}
\begin{document}
\vspace*{0.2in}

\begin{flushleft}
{\Large
\textbf\newline{Modeling Non-Stationarities in High-Frequency Financial Time Series} 
}
\newline
\\
Linda Ponta\textsuperscript{1\Yinyang},
Mailan Trinh\textsuperscript{2\Yinyang},
Marco Raberto\textsuperscript{1\Yinyang},
Enrico Scalas\textsuperscript{2,3\Yinyang,$\ast$},
Silvano Cincotti\textsuperscript{1\Yinyang},
\\
\bigskip
\textbf{1} DIME-CINEF, University of Genoa,  Genoa, Italy
\\
\textbf{2}  Department of Mathematics, School of Mathematical and Physical Sciences, University of Sussex, Brighton, UK
\\
\textbf{3} BCAM - Basque Center for Applied Mathematics, Bilbao, Basque Country - Spain
\\
\bigskip

%
%
\Yinyang These authors contributed equally to this work.





$\ast$ e.scalas@sussex.ac.uk

\end{flushleft}
\section*{Abstract}
We study tick-by-tick financial returns belonging to the FTSE MIB index of the Italian Stock Exchange (Borsa Italiana). We can confirm previously detected non-stationarities. However, scaling properties reported in the previous literature for other high-frequency financial data
are only approximately valid. As a consequence of the empirical analyses, we propose a simple method for describing non-stationary returns, based on a non-homogeneous normal
compound Poisson process. We test this model against the empirical findings and it turns out that the model can approximately reproduce several stylized facts
of high-frequency financial time series. Moreover, using Monte Carlo simulations, we analyze order selection for this model class using three information criteria: Akaike's information criterion (AIC), the Bayesian information criterion (BIC) and the Hannan-Quinn information criterion (HQ).
\textcolor{black}{For comparison, we also perform a similar Monte Carlo experiment for the ACD (autoregressive conditional duration) model. Our results show that the information criteria work best for small parameter numbers for the compound Poisson type models, whereas for the ACD model the model selection procedure does not work well in certain cases.}



\section*{Introduction}
\label{Introduction}
The recent rise in the availability of high-frequency financial data has seen an
increase in the number of studies focusing on the areas of classification and
modeling of financial markets at the ultra-high frequency level. The development of models that are able to reflect
\textcolor{black}{the various phenomena observed in real data} is an important step towards obtaining
a full understanding of the fundamental stochastic processes driving the market.
The statistical
properties of high-frequency financial data and market micro-structural properties were studied by means of different tools,
including phenomenological models of price dynamics
and agent-based market simulations (see
\cite{Goodhart97}, \cite{Ohara99}, \cite{Madhavan00}, \cite{Scalas00}, \cite{Mainardi00}, \cite{Dacorogna01}, \cite{Raberto01}, \cite{Cincotti03}, \cite{Luckock03}, \cite{Scalas04}, \cite{Pastore10}, \cite{Ponta11}, \cite{Ponta11_EPL}, \cite{Ponta12} \cite{Mandelbrot63}, \cite{Mandelbrot97}, \cite{Mueller90}, \cite{Mantegna95}, \cite{Gopikrishnan00}, \cite{Hautsch12}, \cite{Gontis06}, \cite{Gontis07}, \cite{Gontis10}, \cite{Gontis11}, \cite{Kaulakys07}, \cite{Kaulakys07a}, \cite{Kaulakys06}, \cite{Kaulakys06a}, \cite{Kenett13}, \cite{Zheng14}, \cite{Botta15}).

Various studies on high-frequency econometrics appeared in the literature using the
autoregressive conditional duration \textcolor{black}{(ACD)} models (see \cite{Engle97}, \cite{Engle98}, \cite{Bauwens00}, \cite{Lo02}).
Alternative stochastic models were also proposed, e.g., diffusive models, ARCH-GARCH models,
stochastic volatility models, models based on fractional processes, models based on subordinate processes
(see \cite{Cont00}, \cite{Chowdhury99}, \cite{Hardle95}, \cite{Levy95}, \cite{Lux99}, \cite{Stauffer99}, \cite{Youssefmir97}) as well as models based on self-exciting processes of Hawkes type \cite{Hawkes71}, \cite{MuniToke12}, \cite{Sornette15}.
An important variable is the order imbalance. Many existing studies analyze order imbalances around specific events or over
short periods of time. For example, in \cite{Blume89} order imbalances are analyzed around the October 1987
crash. \cite{Chan00} analyzes how order imbalances change the relation
between stock volatility and volume using data for about six months.
A large body of research examines the effect of the bid-ask
spread and the order impact on the short-run behavior of prices (see
\cite{Stoll90}, \cite{Hauser03}, \cite{Chordia02}, \cite{Ponzi09}, \cite{Svorencik07}, \cite{Wyart08},
\cite{Moro09}, \cite{Perello08}, \cite{Preis11}, \cite{Kumaresan10}, \cite{Zaccaria10},
\cite{Lim05}, \cite{Weber05}, \cite{Bouchaud05}).
Trading activity was measured by the average number of trades in unit time intervals in \cite{Bonanno00} and \cite{Plerou00}. However,  aggregating trades into time intervals of the same length may have influences on the analysis. For instance, if intervals are too short with respect to the average waiting time between consecutive trades then every interval will contain either no point or few points. On the contrary, if intervals are too long, aggregation of too many points may lead to loss of information on the time structure of the process. Moreover, in both cases one distorts the kurtosis of the return process (see \cite{Engle98}).

For this reason, \textcolor{black}{an} important empirical variable is the waiting time between two consecutive
transactions (see \cite{Scalas04}, \cite{Scalas06a}, \cite{Gontis08}, \cite{Kaulakys07}, \cite{Gontis06}, \cite{Gontis07}, \cite{Gontis10}, \cite{Gontis11}).
\textcolor{black}{In the market}, during a trading day the activity is not constant (see \cite{Engle97}, \cite{Engle98}) leading to fractal-time behavior
(see \cite{Mandelbrot10}, \cite{Vrobel11}). Indeed, as a consequence of the double auction mechanism, waiting times between two subsequent trades are themselves
\textcolor{black}{random} variables (see \cite{Scalas06a}, \cite{Scalas07}, \cite{Politi08}).
They may also be correlated to returns (see \cite{Raberto02}) as well as to traded volumes.
In the last few years in order to investigate tick-by-tick financial time series, the continuous-time random walk (CTRW)
was used (see \cite{Scalas00}, \cite{Masoliver03}, \cite{Ivanov04}, \cite{Scalas06a}).
It turned out that interorder and intertrade waiting-times are not exponentially distributed. Therefore, the jump process of tick-by-tick prices
is non-Markovian (see \cite{Scalas00}, \cite{Scalas06a}).
Bianco and Grigolini applied a new method to verify whether the intertrade waiting time process is a genuine renewal process
(see \cite{Goldstein04}, \cite{Embrechts97}, \cite{Press92}). This was assumed by the CTRW hypothesis in \cite{Scalas00}.
They found that intertrade waiting-times do follow a renewal process.
Indeed, trading via the order book is asynchronous and a transaction occurs only if a trader issues a market order.
For liquid stocks, waiting times can vary in a range between fractions of a second to a few minutes, depending on the specific
stock and on the market considered. In \cite{Raberto02}, the reader can find a study on General Electric stocks traded in
October 1999. Waiting times between consecutive prices exhibit 1-day periodicity, typical of variable intraday market activity.
Moreover, the survival probability (the complementary cumulative distribution function) of waiting times is not exponentially distributed
(see \cite{Mainardi04}, \cite{Scalas06a}), but is well fitted by a Weibull function (see \cite{Engle94}, \cite{Engle97}, \cite{Engle98}, \cite{Raberto02}, \cite{Ponta12}).

Here, inspired by \cite{Bertram05}, we propose a model based on non-homogeneous Poisson processes.
The paper is organized as follows. Section \ref{SecDataSet} describes the data set. A general description is presented in Subsection
\ref{SubSecGeneralDescription}, and the FTSE MIB index in Subsection \ref{SubSecFTSEMIBIndex}.
Section \ref{SecDescripStat} (and in particular Subsections \ref{SubSecSingleAssets} and \ref{SubSecFTSEMIB}) describes the statistical analysis
of the single assets and of the FTSE MIB index, respectively as well as the scaling analysis; Section \ref{bivariate} contains the bivariate analysis whereas Section \ref{SecModel} is devoted to the compound Poisson model, its order selection and
the numerical results. A comparison with order selection performance for ACD models is presented in the same section. Finally, Section \ref{SecConclusions} presents the conclusions of this work.
\section{Data set}
\label{SecDataSet}
\subsection{General description}
\label{SubSecGeneralDescription}
The data set includes high-frequency trades registered at Italian Stock Exchange (BIt or Borsa Italiana), from the 03$^{\mathrm{rd}}$ of February 2011 to the 09$^{\mathrm{th}}$ of March 2011. The data of February 14$^{\mathrm{th}}$ 2011 are not used because, on that day, there were technical problems at BIt. Moreover, we have removed the data of the 21$^{\mathrm{st}}$ of February, as well. In fact, on that day, there was a crash in the Italian market related to the events in Lybia (on the 15$^{\mathrm{th}}$ of February, a rebellion against the Lybian government begun). We consider the 40 shares in the FTSE MIB index at the time, namely: A2A, STS, ATL, AGL, AZM, BP, BMP, PMI, BUL, BZU, CPR, DIA, ENE, EGP, ENI, EXO, F, FI, FNC, FSA, G, IPG, ISP, LTO, LUX, MS, MB, MED, PLT, PC, PRY, SPM, SRG, STM, TIT, TEN, TRN, TOD, UBI, UCG. Further information on the database and the full meaning of the symbols is available from {\tt www.borsaitaliana.it}. Table \ref{Tab:40AzioniSymbol} shows the meaning of the \textcolor{black}{ticker symbols} as well as the number of observations for each share.
The forty stocks composing the FTSE MIB vary in their average market capitalization and
exhibit different levels of trading activity with different numbers
of trades over this period as summarized in the last column in Table \ref{Tab:40AzioniSymbol} where the total number of observations in the chosen month is given.
Choosing one month of high-frequency data was a trade-off between the necessity of using enough data for significant statistical analysis and, on the other hand, the goal of minimizing the effect of external economic fluctuations leading to non-stationarities of the kind discussed in \cite{Livan12}.
For every stock, the data set consists of prices $p(t_i)$, volumes $v(t_i)$ and times of execution $t_i$, where $i$ is the trade index, varying from $1$ to the total number of daily trades $N$. These data were filtered in order to remove misprints in prices and times of execution. In particular, concerning prices, when there are multiple prices for the same time of execution, we consider only one transaction at that time and a price equal to the average of the multiple prices, and concerning the waiting times, $\tau$, between two executions, we remove observations larger than $200$ $\mathrm{s}$: This means more than $3$ minutes without recorded trading.

\subsection{FTSE MIB Index}
\label{SubSecFTSEMIBIndex}
The FTSE MIB Index (see \cite{FTSEMIBdoc}) is the primary benchmark index for the Italian equity markets. Capturing
approximately 80\% of the domestic market capitalisation, the Index is made up of highly
liquid, leading companies across Industry Classification Benchmark (ICB) sectors in Italy.
The FTSE MIB Index measures the performance of 40 shares listed on Borsa Italiana and
seeks to replicate the broad sector weights of the Italian stock market. The Index is derived
from the universe of stocks trading on BIt. The
Index replaces the previous S\&P/MIB
Index, as a benchmark Index for Exchange Traded Funds (ETFs), and for tracking large
capitalisation stocks in the Italian market. FTSE MIB Index is calculated on a real-time basis in EUR.
The official opening and closing hours of the FTSE MIB Index series coincide with those of
BIt markets and are 09:01 and 17:31 respectively.
The FTSE MIB Index is calculated and published on all days when BIt is open for
trading.

FTSE is responsible for the operation of the FTSE MIB Index. FTSE maintains records of the
market capitalisation of all constituents and other shares and makes changes to the
constituents and their weightings in accordance with the Ground Rules. FTSE carries out
reviews and implement the resulting constituent changes as required by the Ground Rules.
The FTSE MIB Index constituent shares are selected after analysis of the Italian equity universe, to ensure
the Index best represents the Italian equity markets.

The FTSE MIB Index is calculated using a base-weighted aggregate methodology. This means
the level of an Index reflects the total float-adjusted market value of all of the constituent
stocks relative to a particular base period. The total market value of a company is determined
by multiplying the price of its stock by the number of shares in issue (net of treasury shares)
after float adjustment. An indexed number is used to represent the result of this calculation
in order to make the value easier to work with and track over time. As mentioned above, the Index is computed in
real time. The details on how to compute it can be found in \cite{FTSEMIBdoc}.

\section{Descriptive univariate unconditional statistics}
\label{SecDescripStat}
In this section, we separately consider the descriptive univariate unconditional statistics for both the forty assets and for the FTSE MIB Index. By \emph{univariate}, we mean that, here, we do not consider correlations between the variables under study. By \emph{unconditional}, we mean that, here, we do not consider the non-stationary and seasonal behavior of the variables under study and the possible memory effects. Correlation and non-stationarity will be discussed in the next section.
\subsection{Single Assets}
\label{SubSecSingleAssets}
In order to characterize market dynamics on a
trade-by-trade level, we consider three variables: the series of time intervals
between consecutive trades, $\tau$, the trade volumes, $v$, and the trade-by-trade logarithmic returns, $r$.
If $p(t_i)$ represents the price of a stock at time $t_i$ where $t_i$ is the epoch of the $i$-th trade, then we define the return as:
\begin{equation}
r_i = \log \frac{p(t_{i+1})}{p(t_i)}.
\end{equation}
Note that $\tau=t_{i+1}-t_i$ is a random intertrade duration (and not a fixed time interval).

Among the empirical studies on $\tau$, we mention \cite{Golia01,Raberto02}, concerning contemporary shares traded
over a period of a few months, a study on rarely-traded nineteenth century
shares in \cite{Sabatelli02}, and results on foreign exchange transactions in \cite{Takayasu02} and \cite{Marinelli01}.

Tables  \ref{Tab:StatisticaTempiAttesa}, \ref{Tab:StatisticaVolumi} and \ref{Tab:StatisticaPrice} contain the descriptive statistics, evaluated for the entire sample, for the time series $\tau^h_i=t^h_{i+1}-t^h_{i}$ (with $t^h_0=0$), $v^h_i$ and $r^h_i$, where the superscript $h$ denotes the specific share.

\begin{figure*}[h]
\centering
\includegraphics[width=4.5in]{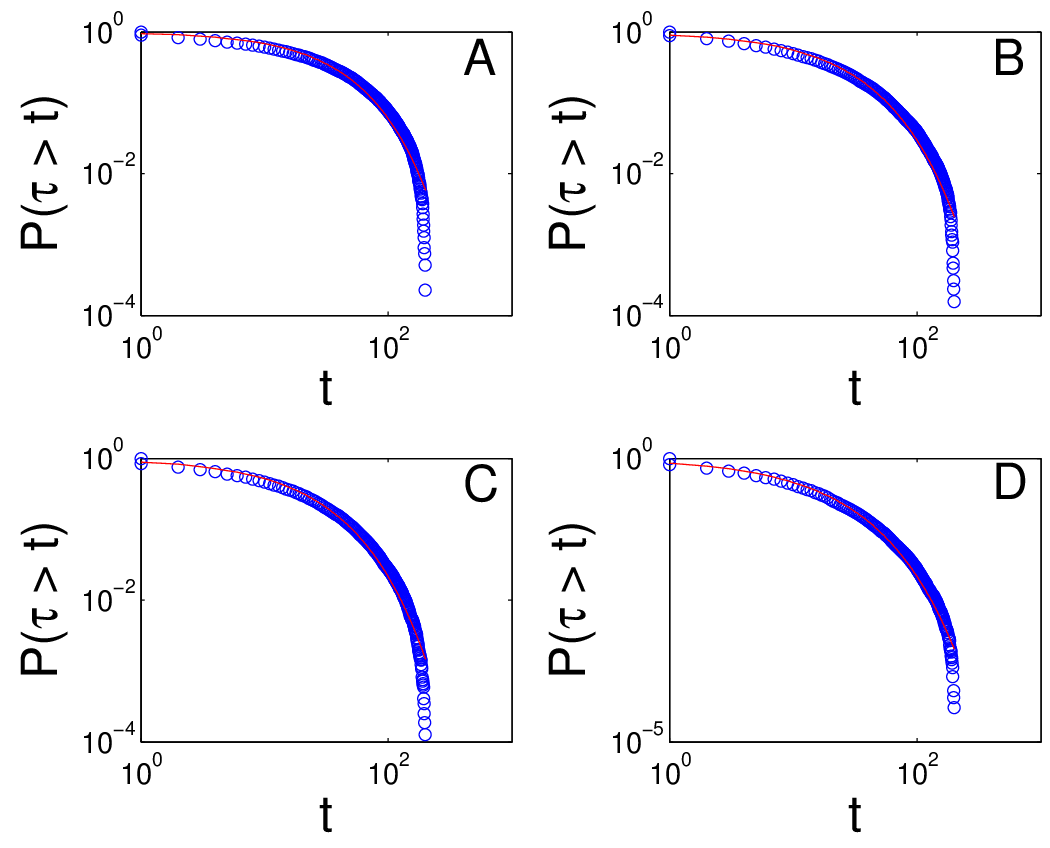}
\caption{Weibull fit for A2A (A), EXO (B), MS (C), TIT (D). The fit is represented by the thin solid line, the open circles are the empirical values for the survival function.}
\label{Fig:FitWeibullAsset}
\end{figure*}

\begin{figure*}[h]
\centering
\includegraphics[width=3.5in]{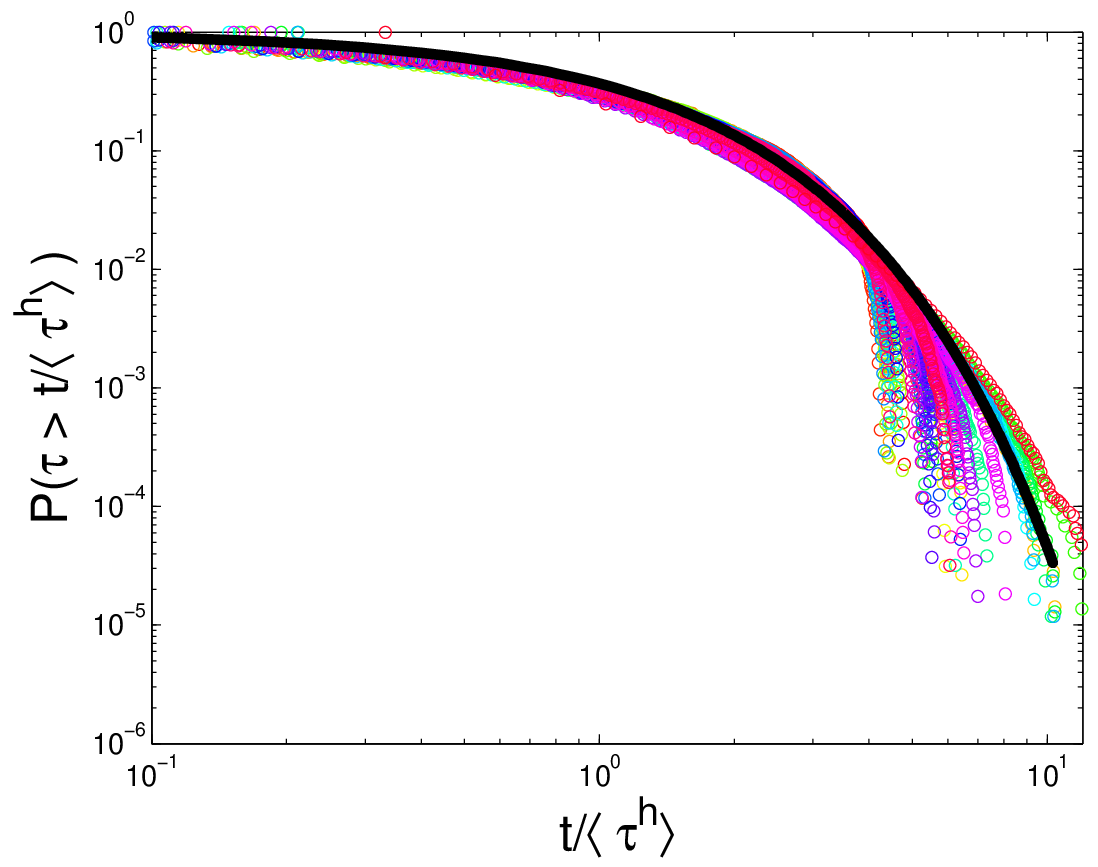}
\caption{Approximate scaling of the survival function for the forty time series. The solid line is the Weibull fit given by Eq.\eqref{Eq:WeibScaling}.}
\label{Fig:FitWeibullAssetTot}
\end{figure*}

\begin{figure*}[h]
\centering
\includegraphics[width=2.5in]{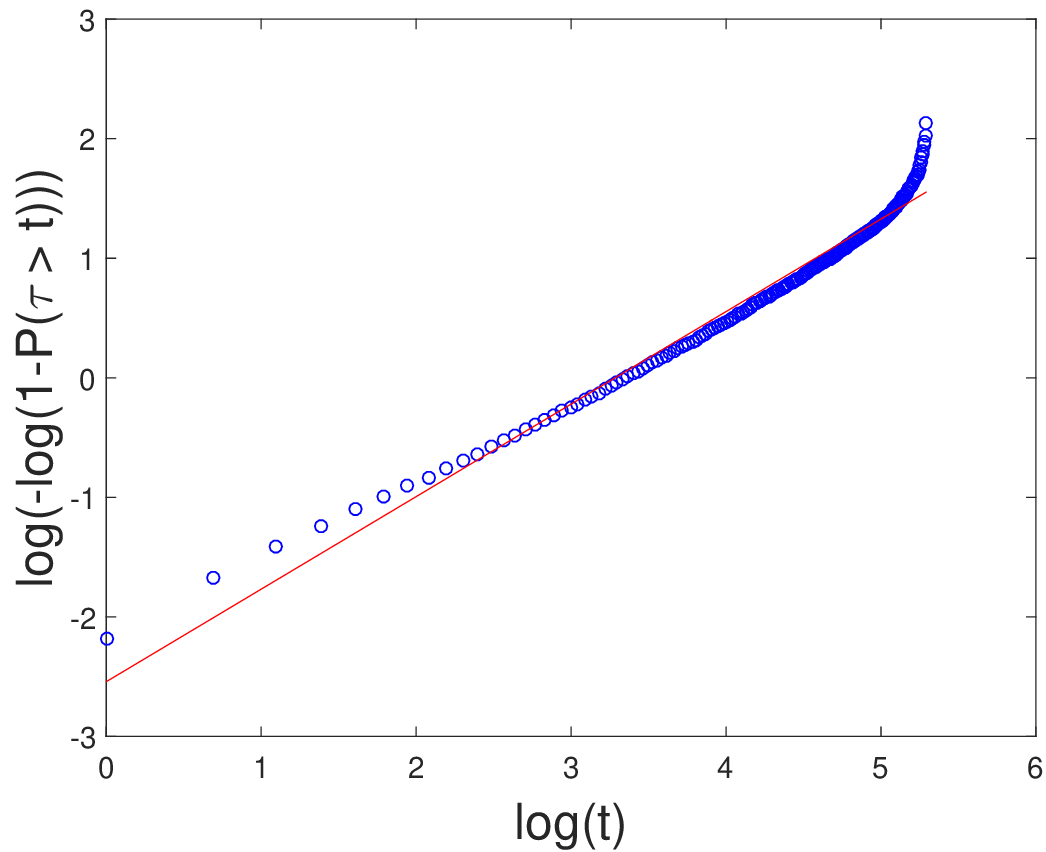}
\includegraphics[width=2.5in]{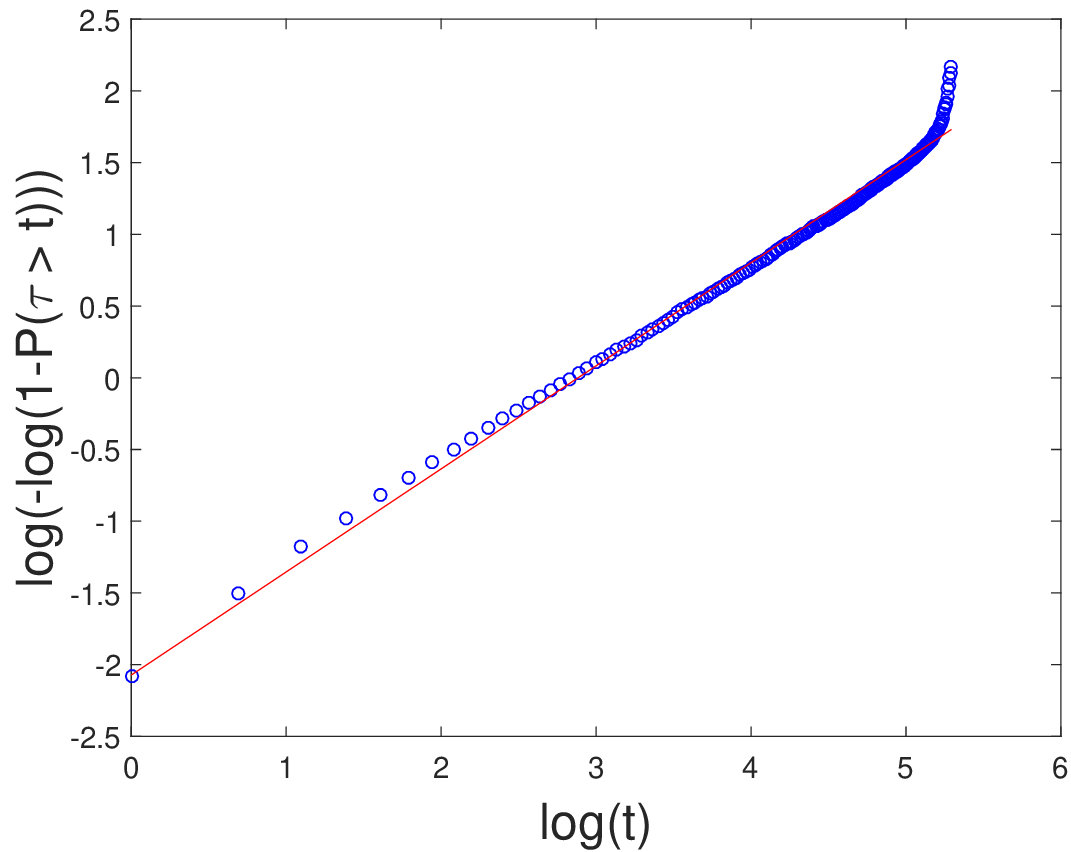}
\includegraphics[width=2.5in]{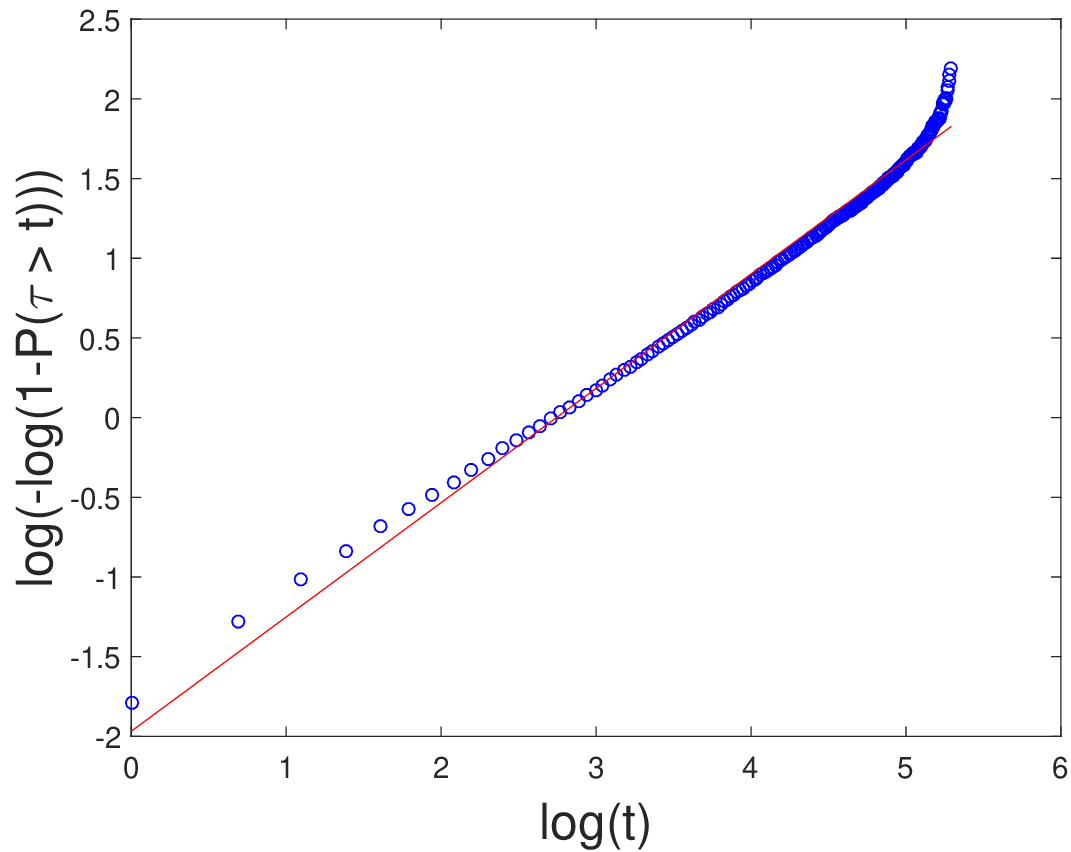}
\includegraphics[width=2.5in]{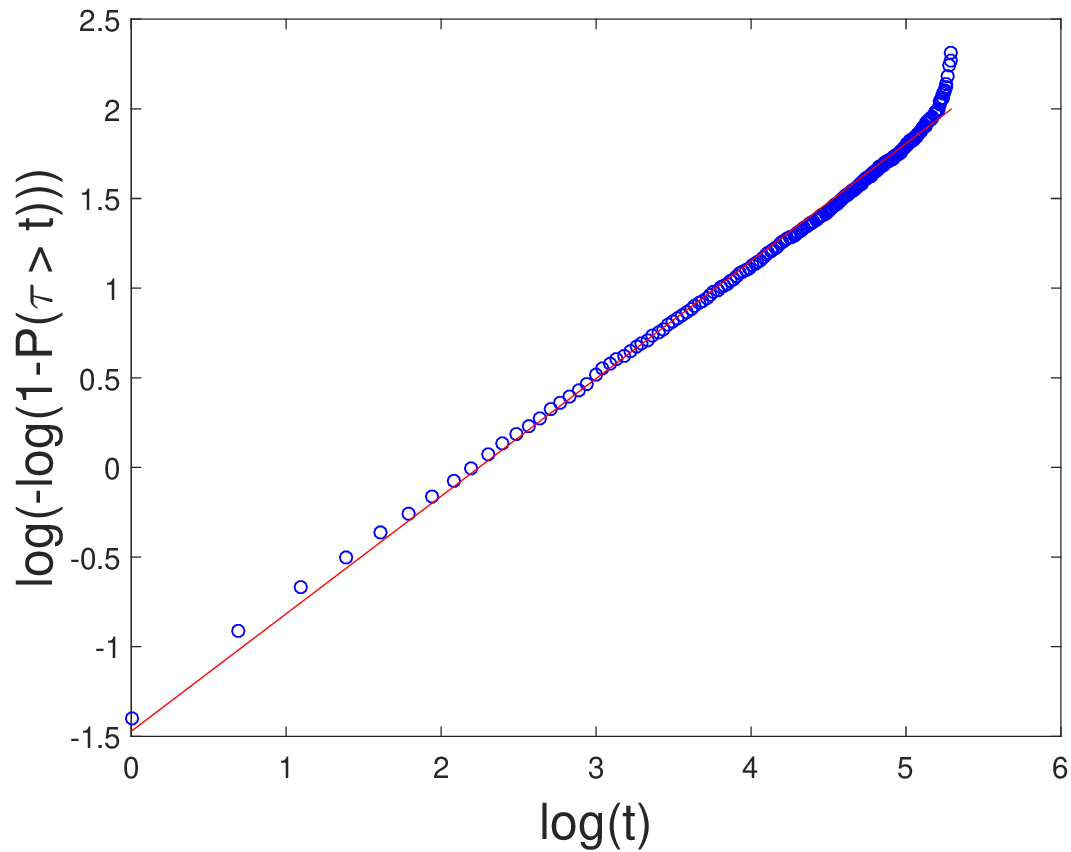}
\caption{Weibull paper for A2A (A), EXO (B), MS (C), TIT (D). On the horizontal axis, the values of $\log(t)$ are plotted, where $t$ represents the inter-trade duration. On the vertical axes, a double logarithmic transform of the empirical cumulative distribution function of the inter-trade durations is plotted: $\log(-\log(1-P(\tau> t)))$.  The linear fit is represented by the thin red solid line, the open circles are the empirical values.}
\label{Fig:FitWeibullAssetPaper}
\end{figure*}

\begin{figure*}[h]
\centering
\includegraphics[width=4.5in]{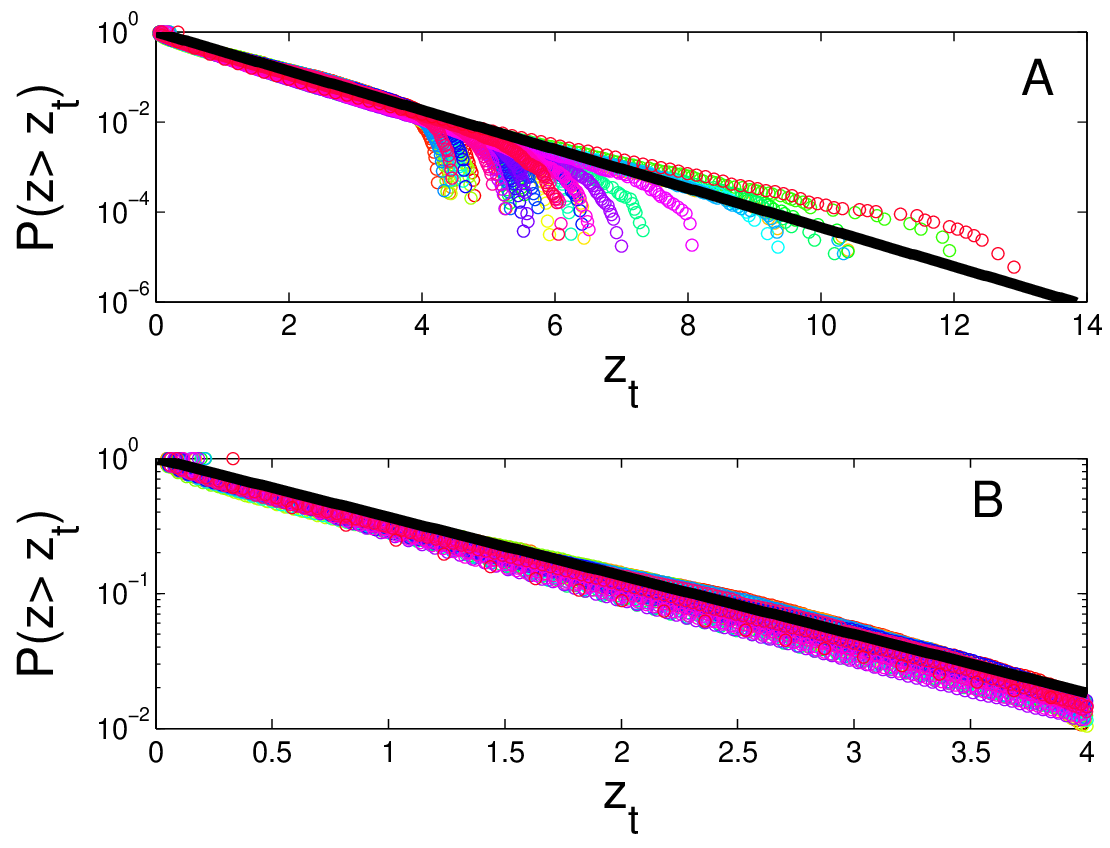}
\caption{(A) Empirical survival function for the transformed variable $z_t^h$ compared with the expected exponential function $\exp(-z_t^h)$; (B) Zoom in the region $z_t^h \leq 4$. }
\label{Fig:FitWeibulEXPZoom}
\end{figure*}

In Table \ref{Tab:StatisticaTempiAttesa} the third and fourth columns give the two parameters of a Weibull distribution fit. The Weibull distribution has the following survival function:
\begin{equation}
\mathbb{P}(\tau>t)=P(t|\alpha,\beta)=\exp\left(-\alpha t^\beta\right),
\end{equation}
where $\beta$ is the shape parameter and $\alpha$ is the scale parameter.
The values given in Table \ref{Tab:StatisticaTempiAttesa} were fitted using the moment method described in \cite{Politi08}.
The quality of these fits is pictorially shown in Fig \ref{Fig:FitWeibullAsset} for A2A, EXO, MS and TIT, respectively. The solid line represents our Weibull fit and the circles are the empirical data. Since different companies have different average intertrade
duration $\langle \tau^h \rangle$ (see the second column in Table \ref{Tab:StatisticaTempiAttesa}), they are also characterized by
a different scale parameter $\alpha$ whereas the shape parameter $\beta$ is almost the same for all the forty time series. Following \cite{Ivanov04}, a scaling function $P(t|\beta^*)$ can be defined:
\begin{equation}
\label{Eq:WeibScaling}
P(t|\beta^*)=\exp\left(- (t/\langle \tau \rangle)^{\beta^*}\right)
\end{equation}
where $\beta^*= \langle \beta \rangle = 0.78$.
To test the hypothesis that there is a universal structure in the intertrade time dynamics of different companies,
we rescale the survival functions by plotting them against $t/ \langle \tau^h \rangle$. We find that, for all companies,
data approximately conform to a single scaled plot given by \eqref{Eq:WeibScaling} as shown in Fig \ref{Fig:FitWeibullAssetTot} (see also \cite{Stauffer96,Ivanov04,Politi08}). Such a behavior is a hallmark of scaling, and
is typical of a wide class of physical systems with universal
scaling properties \cite{Bunde94}.
Even if \cite{Eisler06} showed that the scaling \eqref{Eq:WeibScaling} is far from being universal, at least for the New York Stock Exchange, it is remarkable to find it again for a different index in a different market and seven years later with respect to the findings of \cite{Ivanov04}. However, to go beyond qualitative estimates, the goodness-of-fit test is given in the sixth column of Table \ref{Tab:StatisticaTempiAttesa}, where we report the Anderson-Darling (AD) statistics for the transformed random variable $z_{\tau}^h= \alpha \tau^{\beta}$. $z_{\tau}$ should follow an exponential distribution with parameter $\mu=1$, if $\tau$ is distributed according to a Weibull distribution. A glance at Fig \ref{Fig:FitWeibulEXPZoom} immediately shows that this is not the case; for $z_{\tau}>4$ there are significant deviations from the exponential law, whereas this is approximately satisfied for $z_{\tau}\leq 4$. This fact is reflected by the high values of the AD statistics for which the critical value at $0.05$ significance level is 1.34. In other words, the Weibull null hypothesis is rejected for all the time series.
\textcolor{black}{To confirm these results, we also perform the Lilliefors test. The Lilliefors statistics are larger that the critical value at $0.05$ significance level as well. Furthermore, we peform the Kolmogorov-Smirnov test in order to check if the distributions in Fig \ref{Fig:FitWeibulEXPZoom} collapse into one. The results show that the null hypothesis of same distribution is always rejected (contact the authors for full tables).
Then, we perform the AD test and the Lilliefors test for the index durations, and also in this case the null hypothesis of Weibull distribution is rejected by both statistical tests.
Finally we present results based on the Weibull paper to graphically verify the Weibull distribution hypothesis. As an illustration, Fig \ref{Fig:FitWeibullAssetPaper} shows the Weibull paper for the following assets: A2A, EXO, MS and TIT. We can see the deviation of the empirical data from the straight line expected for the Weibull distribution.}

In this paper, we do not study volumes $v^h$, but we \textcolor{black}{present} their descriptive statistics in Table \ref{Tab:StatisticaVolumi} for the  sake of completeness.

\begin{figure*}[h]
\centering
\includegraphics[width=4.5in]{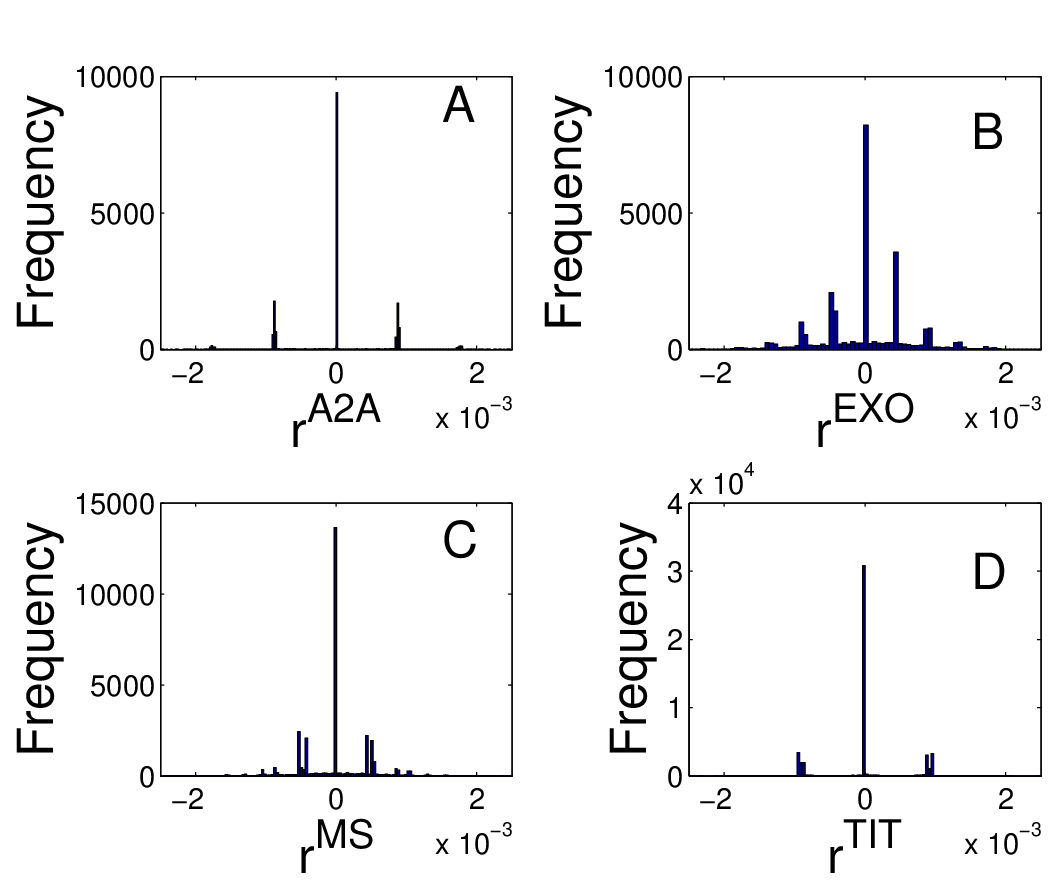}
\caption{Histogram of returns for A2A (A), EXO (B), MS (C), TIT (D).}
\label{Fig:HistAsset}
\end{figure*}

The descriptive statistics for trade-by-trade returns $r^h$ can be found in Table \ref{Tab:StatisticaPrice}. Notice that there is excess kurtosis.
The histograms in Fig \ref{Fig:HistAsset} for the asset prices A2A, EXO, MS and TIT, respectively, show how the returns are distributed. It is possible to appreciate the discrete character of returns even after the logarithmic transformation.

\subsection{FTSE MIB index}
\label{SubSecFTSEMIB}
As well as the single assets, we investigate the FTSE MIB index.
Tables \ref{Tab:StatisticaTempiAttesa} and \ref{Tab:StatisticaPrice}  summarize also the descriptive statistics of the time series $\tau^I_i$ and $r_i^I$ respectively evaluated for the FTSE MIB index as trade-by-trade volumes are not available.

\begin{figure*}[h]
\centering
\includegraphics[width=3.5in]{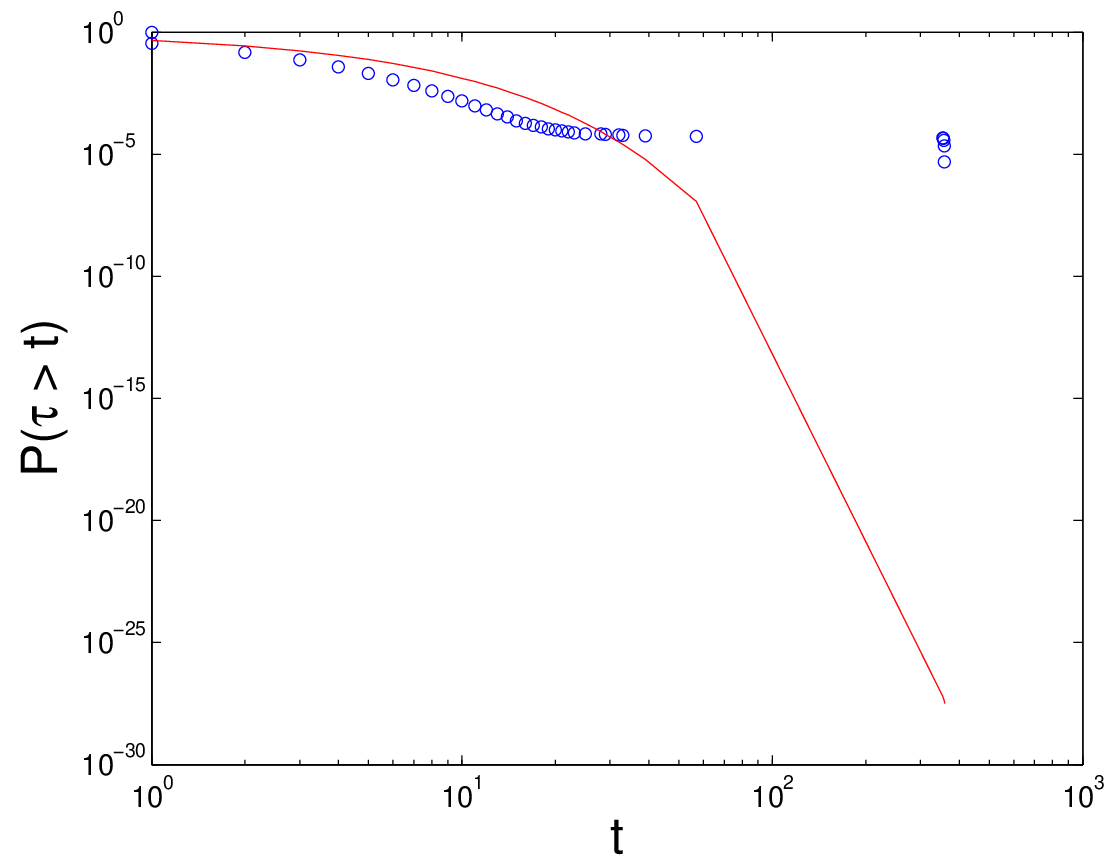}
\caption{Circles: empirical survival function; solid line: Weibull fit.}
\label{Fig:FitWeibullIndice}
\end{figure*}

In Fig \ref{Fig:FitWeibullIndice} we show the survival function for the intertrade waiting time of the FTSE MIB index. The solid line represents the Weibull fit, whereas the circle represents the empirical data. The shape of the two curves is very different. Therefore, we can immediately see that intertrade times are not Weibull distributed, and, in this case, the fit does not work even as a first approximation. Indeed, for the FTSE MIB index, the standard deviation of intertrade durations is smaller than the average intertrade duration and the AD test and the Lilliefors test reject the null hypothesis of Weibull distribution as discussed previously.

\begin{figure*}[h]
\centering
\includegraphics[width=3.5in]{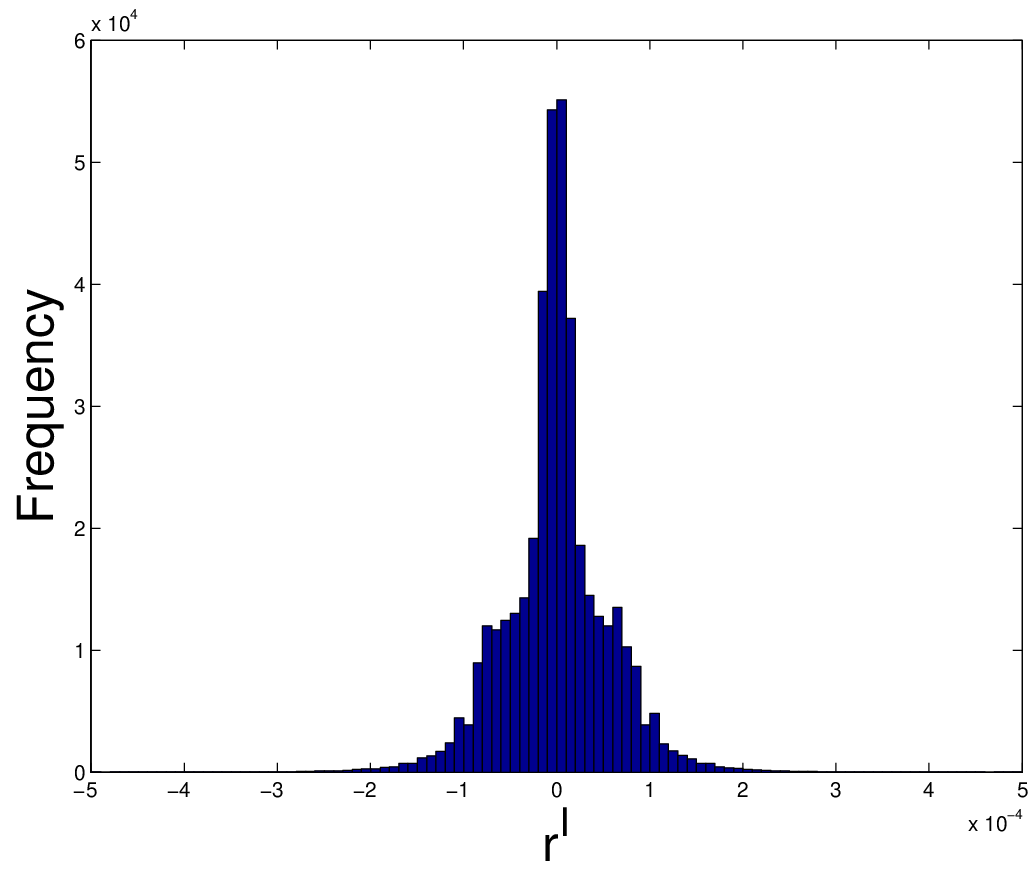}
\caption{Histogram of returns for the FTSE MIB index.}
\label{Fig:HistIndice}
\end{figure*}

Contrary to the case of single asset returns, the excess kurtosis for the FTSE MIB index is quite large.
Fig \ref{Fig:HistIndice} shows the histogram of the returns for a bin size of $1 \times 10^{-5}$.

\begin{figure*}[h]
\centering
\includegraphics[width=4.5in]{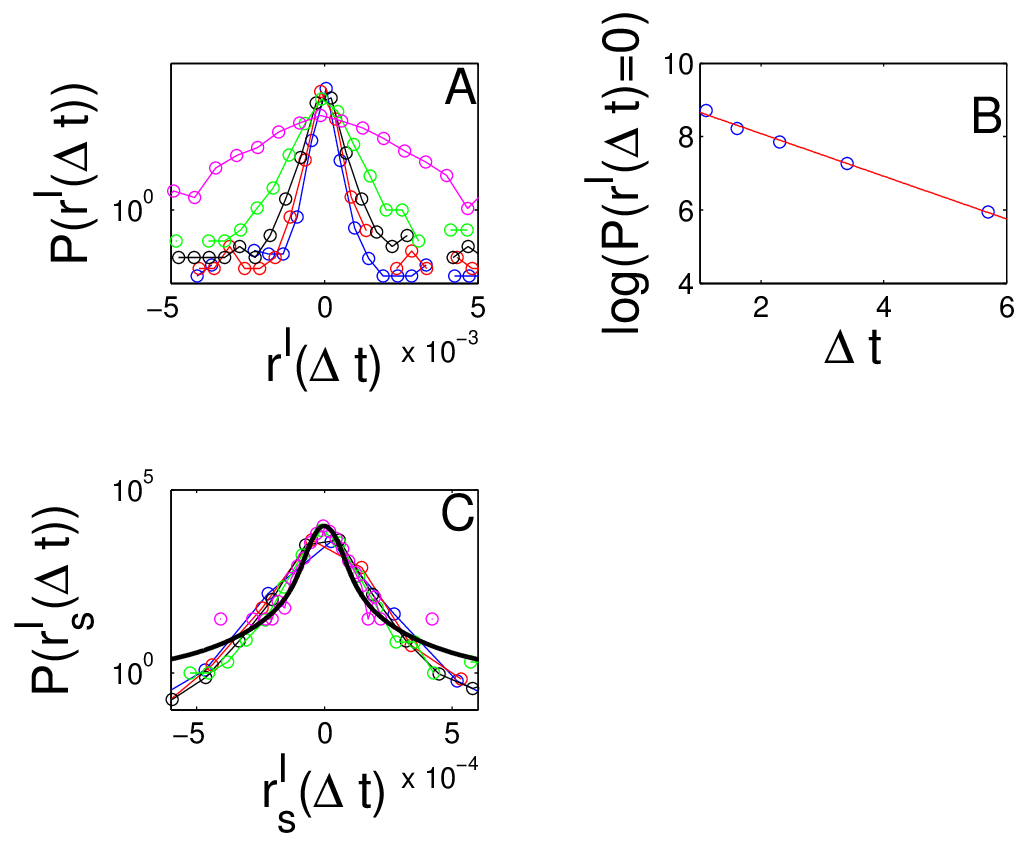}
\caption{(A) Histogram of the returns for the FTSE MIB index observed at different time intervals, namely, $\Delta t=$ $3$ s (blue),  $5$ s (red), $10$ s (black), $30$ s (green) and $300$ s (purple); (B) Probability of zero returns as a function of the time sampling interval $\Delta t$, the slope of the straight line is $0.58\pm 0.01$; (C) scaled empirical probability distribution and comparison with the theoretical prediction  given by Eq.\eqref{Eq:scaling2} (black solid line).}
\label{Fig:Scaling}
\end{figure*}

Following \cite{Mantegna95}, we test the scaling of the empirical returns.
As shown in Table \ref{Tab:40AzioniSymbol}, the dataset consists of 405560 records for the FTSE MIB index during the period studied (from the 03$^{\mathrm{rd}}$ of February 2011 to the 09$^{\mathrm{th}}$ of March 2011). From this database, we compute the new random variable $r^I(t; \Delta t)$ defined as:
\begin{equation}
\label{Eq:SampledReturns}
r^I(t; \Delta t)=\log \frac{p^I(t+ \Delta t)}{p^I(t)},
\end{equation}
where $p^I(t)$ is the value of the index at time $t$.
In this way we sample returns on equally spaced and non-overlapping intervals of width $\Delta t$. We further assume that the time series is stationary so that it only depends on $\Delta t$ and not on $t$ (incidentally, we shall see that this is not the case). To characterize quantitatively the experimentally observed process, we first determine the empirical probability density function $P(r^I(\Delta t))$ of index variations for different values of $\Delta t$. We select $\Delta t$ equal to 3s, 5s, 10s, 30s and 300s.
In Fig \ref{Fig:Scaling}(a) we present a semi-logarithmic plot of $P(r^I(\Delta t))$ for the five different values of $\Delta t$ indicated above. These empirical distributions are roughly symmetric and are expected to converge to the normal distribution when $\Delta t$ increases.
\textcolor{black}{The null hypothesis of normal distribution has been tested with the Kolmogorov-Smirnov, the Jarque-Bera and the Lilliefors test. The results reported in Table \ref{Fig7a} show that the null hypothesis is always rejected.}

We also note that the distributions are leptokurtic, that is, they have tails heavier than expected for a normal process. A determination of the parameters characterizing the distributions is difficult especially because larger values of $\Delta t$ imply a smaller number of data.
Again following \cite{Mantegna95}, we study the probability density at zero return $P(r^I(\Delta t)=0)$ as function of $\Delta t$. This is done in Fig  \ref{Fig:Scaling}(b), where $P(r^I(\Delta t)=0)$ versus $\Delta t$ is shown in a log-log plot.
If these data were distributed according to a symmetric $\alpha$-stable distribution, one would expect the following form for $P(r^I(\Delta t)=0)$:
\begin{equation}
\label{Eq:Po}
P(r^I(\Delta t)=0)=\frac{\Gamma(1/\alpha_L)}{\pi \alpha_L (c \Delta t)^{1/\alpha_L}},
\end{equation}
where $\Gamma(\cdot)$ is Euler Gamma function, $\alpha_L \in (0,2]$ is the index of the symmetric $\alpha$-stable distribution and $c$ is a time-scale parameter.
The data are well fitted (in the OLS sense) by a straight line of slope $1/\hat{\alpha}_L=0.58$ leading to an estimated exponent $\hat{\alpha}_L=1.72$. The best method to get the values of $P(r^I(\Delta t)=0)$ is to determine the slope of the cumulative distribution function in $r^I(\Delta t)=0$.
In Fig \ref{Fig:Scaling}(c), we plot the rescaled probability density function according to the following transformation:
\begin{equation}
\label{Eq:scaling1}
r^I_s=\frac{r^I(\Delta t)}{(\Delta t)^{1/\alpha_L}}
\end{equation}
and
\begin{equation}
\label{Eq:scaling2}
P(r^I_s)=\frac{P(r^I(\Delta t))}{(\Delta t)^{-1/\alpha_L}},
\end{equation}
for $\alpha_L=\hat{\alpha}_L=1.72$. Remarkably all the five distributions approximately collapse into a single one. \textcolor{black}{We use the Kolmogorov-Smirnov test to study the null hypothesis of identically distributed rescaled data; the results are shown in Table \ref{Tab:Fig7c}. The null hypothesis is rejected only in the following cases: $\Delta t = 3s$ and $\Delta t = 5s$, $\Delta t = 3s$ and $\Delta t = 10s$, $\Delta t = 3s$ and $\Delta t = 30s$.  }
It is worth noting that this result shows that the scaling, found in the S\&P 500 data by Mantegna and Stanley more than twenty years ago \cite{Mantegna95}, still approximately holds in a different market and in a completely different period. We do not run hypothesis tests on the L\'evy stable distribution because an eye inspection of Fig \ref{Fig:Scaling}(c) is sufficient to conclude that the L\'evy stable fit is not matching the rescaled data.

\section{Descriptive conditional and bivariate statistics}
\label{bivariate}
\begin{figure*}[h]
\centering
\includegraphics[width=4.5in]{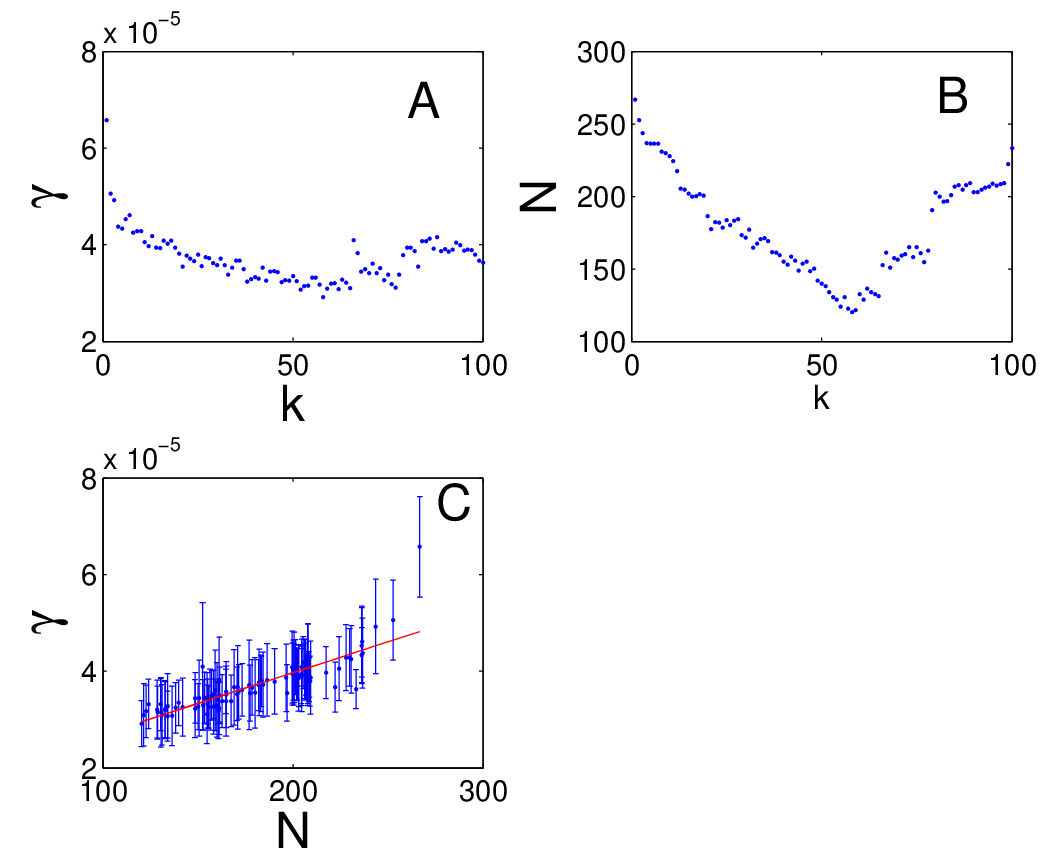}
\caption{(A) Volatility $\gamma$ as a function of $k$ for $\delta t=300$ s. (B) Activity $N$ as a function of $k$  for $\delta t=300$ s. (C) Scatter plot of volatility $\gamma$ as a function of number of trades $N$. The points are averaged over the investigated period.}
\label{Fig:GammaNgDt300}
\end{figure*}

Inspired by \cite{Bertram04,Bertram05}, in order to study the time variations of the returns during a typical trading day, we use a simple technique. We divide the trading day into equally spaced and non-overlapping intervals of length $\delta t$ for $\delta t= 3, 5, 10, 30, 300, 600, 900, 1200, 1500$ and $1800$ s.
Let $K$ be number of intervals and $N_k$ the number of transaction in each interval $k$. For each interval we evaluate the $\gamma(k)$ indicator as a measure of volatility.
$\gamma(k)$ is defined as
\begin{equation}
\gamma(k) = \frac{1}{N_k-1}\sum_{i=1}^{N_k-1}|r^I_{k,i}-\langle r^I_k \rangle|;
\end{equation}
where $\langle r^I_k \rangle$ is the average value of returns in the time interval $k$. In Fig \ref{Fig:GammaNgDt300}(a), as an example, we plot the average value of $\gamma(k)$ over the investigated period as a function of the interval index $k$ for $\delta t= 300 $ s. We can see that the volatility is higher in the morning, at the opening of continuous trading, and then it decreases up to midday. There is a local increase after midday and then the volatility returns to lower values to finally grow towards the end of continuous trading.
The above pattern can be reinforced by the presence of the many \emph{day traders}
whose practice is to close all their positions at the end of each trading day and
reopen them in the following morning. The rationale of day traders is to avoid
overnight exposure to risk. Interestingly, this plot also provides us with a picture of
the social behavior of Borsa Italiana equity traders.
The volatility can be seen to drop off in the interval $12$:$30$ - $14$:$00$ and to grow suddenly again around $14$:$20$. These times correspond to the typically preferred lunchtime interval of most traders.
In Fig \ref{Fig:GammaNgDt300}(b), we plot the number of trades on the FTSE MIB index as a function of the interval index $k$ for $\delta t=300$ s. The behavior of the trade activity closely follows the behavior of volatility. This is even clearer from the analysis of Fig \ref{Fig:GammaNgDt300}(c) where the volatility is plotted as a function of the activity. The scatter plot shows a strong correlation between the two variables.
This result does not depend on the length of the interval $w$, but the corresponding plots are not presented here for the sake of compactness.
This feature was already present in the Australian market studied for a much longer period (10 years $ \approx$ 2500 days) by \cite{Bertram04,Bertram05}. Again, it is remarkable to see a statistical pattern still valid in a different market after more than 10 years.

Fig \ref{Fig:GammaNgDt300} shows a clear seasonal pattern in intraday trades. In order to take this behavior into account, we proposed to use a non-stationary normal compound Poisson process with volatility of jumps proportional to the activity of the Poisson process in \cite{Scalas07}.
Here, we take even a more pragmatic stand and we do not assume any \emph{a priori} relationship between volatility and activity as it emerges spontaneously, if present, with the method described in the next section.

\begin{figure*}[h]
\centering
\includegraphics[width=3.5in]{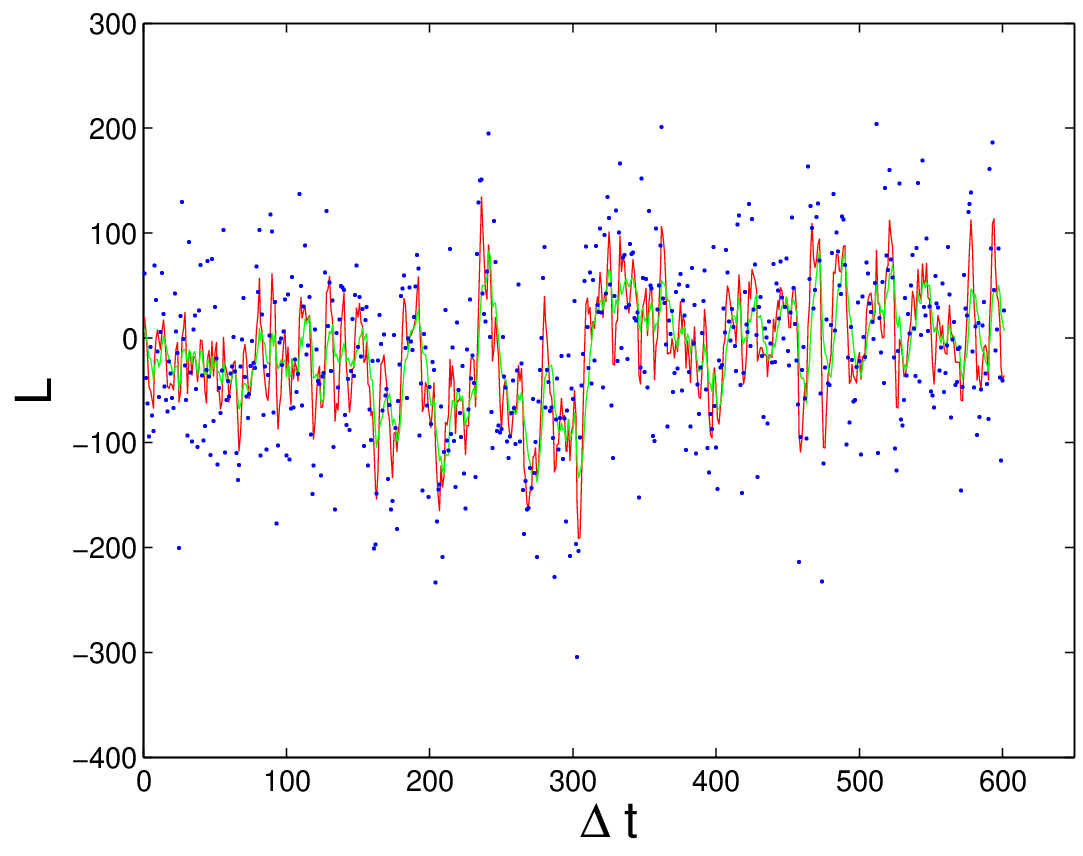}
\caption{Leverage $L$ as a function of lag $\Delta t$. The red and green solid lines show the leading short ($4$ lags) and lagging long ($10$ lags) square-root weighted moving average, respectively. $\Delta t$ is equal to $3$s. There is no strong evidence of leverage effect.
}
\label{Fig:LeverageDR}
\end{figure*}

\begin{figure*}[h]
\centering
\includegraphics[width=4.5in]{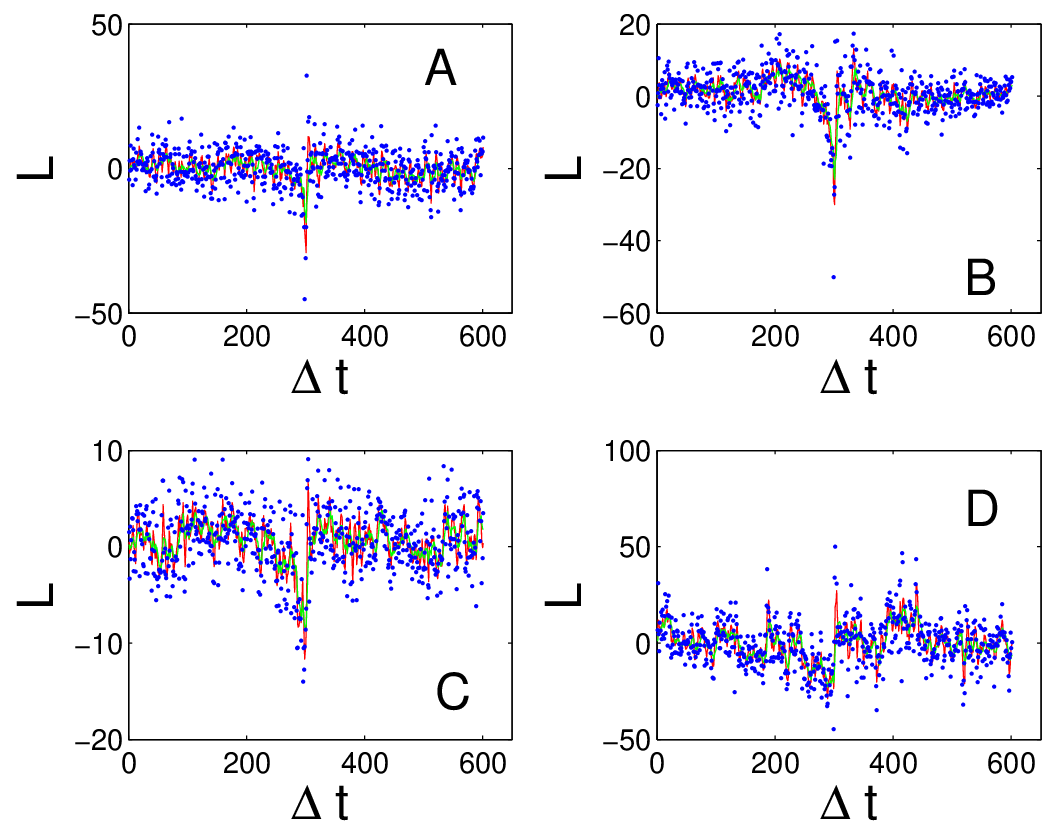}
\caption{ Leverage $L$ as a function of lag $\Delta t$ for S\&P500 index (A), NASDAQ index (B), CAC40 index (C), FTSE index (D). The red and green solid lines show the leading short ($4$ lags) and lagging long ($10$ lags) square-root weighted moving average, respectively. The $\Delta t$ is equal to one day. For S\&P500, NASDAQ indices, the leverage effect is well evident, whereas for CAC40 and FTSE indices it is less evident.
}
\label{Fig:Leverage_SNCF}
\end{figure*}

\begin{figure*}[h]
\centering
\includegraphics[width=4.5in]{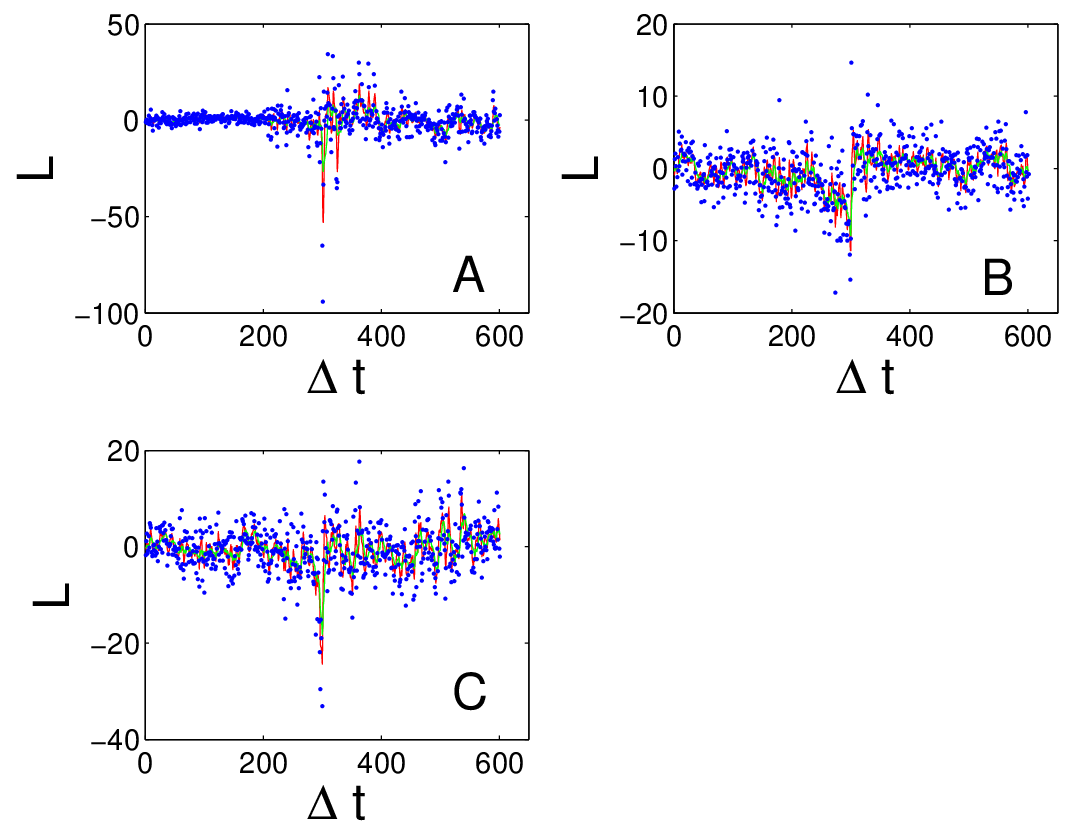}
\caption{ Leverage $L$ as a function of lag $\Delta t$ for DAX index (A), Nikkei index (B), Hang Seng index (C). The red and green solid lines show the leading short ($4$ lags) and lagging long ($10$ lags) square-root weighted moving average, respectively. The $\Delta t$ is equal to one day. For DAX, Nikkei and Hang Seng indices, the leverage effect is well evident.}
\label{Fig:Leverage_DNH}
\end{figure*}

Empirical studies of volatility for daily financial data by \cite{Bouchaud01} have shown that volatility estimates and returns are negatively correlated for positive time lag. Therefore, following \cite{Bouchaud01}, we investigate
this effect on high frequency data by estimating the leverage correlation function as
\begin{equation}
L(\Delta t) = \frac{\langle(r^{I}(t+\Delta t))^2r^I(t)\rangle}{(\mathrm{var}[r^I(t)])^2},
\end{equation}
where $\Delta t$ represents the lag.
The estimates for empirical data samples are shown in Fig \ref{Fig:LeverageDR} for $\Delta t = 3$s. The leverage effect is not evident. For comparison, in Figs \ref{Fig:Leverage_SNCF} and \ref{Fig:Leverage_DNH}, we computed $L(\Delta t)$ for $7$ major international stock indices (S\&P500, NASDAQ, CAC40, FTSE, DAX, Nikkei, Hang Seng) for $\Delta t$ equal to one day. The dataset consisted of daily close prices adjusted for dividends and splits ranging from January 1990 to October 2000 as in \cite{Bouchaud01}.
In the case of S\&P500, NASDAQ, DAX, Nikkei and Hang Seng indices, the leverage effect is well evident, whereas for CAC40 and FTSE indices it is less evident. However, in all these cases, the leverage effect is much stronger then in our high frequency data, if any.

\section{A compound Poisson type model}
\label{SecModel}

As one can see, during a trading day, the volatility and the activity are higher at the opening of the market, then they decrease at midday and they increase again towards market closure \cite{Bertram04} (see also Fig \ref{Fig:GammaNgDt300}). In other words, the (log-)price process is non-stationary. As suggested in \cite{Scalas07}, such a non-stationary process for log-prices can be approximated by a mixture of normal compound Poisson processes (NCPP) in the following way. A normal compound Poisson process is a compound Poisson process with normal jumps. In formula:
\begin{equation}
\label{Eq:NCPP}
X(t)=\sum_{i=1}^{N(t)} R_i,
\end{equation}
where $R_i$ are normally distributed independent trade-by-trade log-returns, $N(t)$ is a Poisson process with parameter $\lambda$ and $X(t)$ is the logarithmic price, $X(t)=\log(P(t))$.
By probabilistic arguments one can derive the cumulative distribution function of $X(t)$, it is given by:
\begin{equation}
\label{Eq:NCPPCDF}
F_{X(t)}(u)=\mathbb{P}(X(t)\leq u)=\mathrm{e}^{-\lambda t}\sum_{n=0}^{\infty}\frac{(\lambda t)^n}{n!}F^{\star n}_R(u),
\end{equation}
where $F^{\star n}_R(u)$ is the $n$-fold convolution of the normal distribution, namely
\begin{equation}
F^{\star n}_R(u)=\frac{1}{2}\left[ 1+ \mathrm{erf}\left(\frac{u-n \mu}{\sqrt{2n \sigma^2}}\right) \right],
\end{equation}
and $\mu$ and $\sigma^2$ are the parameters of the normal distribution.

We now assume that the trading day can be divided into $n$ equal intervals of constant activity $\{\lambda_i\}^n_{i=1}$ and of length $w$, then the unconditional waiting time distribution becomes a mixture of exponential distributions and its cumulative distribution function can be written as
\begin{equation}
\label{Eq:Mistura}
F_{\tau}(u)=\mathbb{P}(\tau\leq u)=\sum_{i=1}^n a_i (1- \mathrm{e}^{-\lambda_i \tau}),
\end{equation}
where $\{a_i\}_{i=1}^n$ is a set of suitable weights. The activity seasonality can be mimicked by values of \textcolor{black}{$\lambda_i$} that decrease towards midday and then increase again towards market closure. In order to reproduce the correlation between volatility and activity, one could assume that
\begin{equation}
\sigma_{\xi,i}=c \lambda_i
\end{equation}
where $c$ is a suitable constant. As already mentioned, however, for practical purposes, one can also estimate three parameters for each interval, the parameter $\lambda_i$ of the Poisson process and the parameters $\mu_i$ and $\sigma_i$ for the log-returns without any correlation assumptions. This leads us to two possible examples of such compound Poisson type models which will be introduced in Section \ref{sec:definition} alongside the popular ACD model for later comparisons. After a brief error analysis of the maximum likelihood estimation (MLE) method in Section \ref{sec:mle-goodness-fit} we will move on to the main Monte Carlo experiment to test model selection using information criteria (IC) in Section \ref{sec:model-selection}. The different nature of the compound Poisson models and the ACD model makes a direct comparison in terms of model selection questionable. Therefore, our main focus will be a comparison of IC within each model class separately.

\subsection{Model definitions and likelihood functions}
\label{sec:definition}

\subsubsection{The compound Poisson model with discrete intensity (D$\lambda$)-model}
\label{sec:comp-poiss-model}
We extend the notation of Eq (\ref{Eq:NCPP}) by an additional index denoting the corresponding interval: We suppose that high-frequency data is given over a time interval $[t_0, T]$. First, set a time grid $\{t_i\}_{i \in \{1, \ldots, n\}}$ such that $t_0 < t_1 < t_2 < \ldots < t_n=T$.
On each time interval $[t_{i-1},t_i)$ we have a compound Poisson process
\begin{equation}
  \label{eq:10}
  X_i(t) := \sum_{k=1}^{N_i{(t)}} R_k^{(i)},
\end{equation}
where $\{R_k^{(i)}\}_{k \in \mathbb{N}}$ is an i.i.d. sequence of  $\mathcal{N}(\mu_i, \sigma_i^2)$ distributed random variables and $(N_i(t))_{t \geq 0}$ is a homogeneous Poisson process with parameter $\lambda_i$. Further, $\{R_k^{(i)}\}_{k \in \mathbb{N}}$ are all independent of $(N_i(t))_{t \geq 0}$.\\
For a fixed time interval $[t_{i-1}, t_i)$ the log-likelihood function is given by
\begin{equation}
  \label{eq:12}
  \mathcal{L}^D_i(\lambda_i, \mu_i, \sigma_i) = -\lambda_i (t_{i}-t_{i-1}) + \ln(\lambda_i) N_i(t_i) + \sum_{k=1}^{N_i(t_i)} \ln(p_{\mu_i,\sigma_i}(R_k^{(i)})),
\end{equation}
where $p_{\mu_i, \sigma_i}$ denotes the probability density function of the $\mathcal{N}(\mu_i, \sigma_i^2)$ distribution. Due to the independence assumptions the overall log-likelihood is given by the sum of all $\mathcal{L}_i$. Eq (\ref{eq:12}) can be derived from the general expression for the sample density function given on p. 200 in \cite{snyder} by substituting a constant $\lambda$.\\
The maximum likelihood estimators are therefore:
\begin{equation}
  \hat{\lambda}_i = N_i/w_i; \qquad \hat{\mu}_i = \frac{1}{N_i}\sum_{k=1}^{N_i} r_i; \qquad \hat{\sigma}^2_i = \frac{1}{N_i}\sum_{k=1}^{N_i} (r_i-\hat{\mu}_i)^2, \label{eq:2}
\end{equation}
where $N_i$ is the number of trades in the $i$th interval and \textcolor{black}{$w_i = t_i - t_{i-1}$.}\\
Note that the maximum likelihood estimator for $\sigma^2$ is biased and the bias can be corrected by using
\begin{equation}
  \label{eq:1}
  \tilde{\sigma}^2_i= \frac{1}{N_i-1}\sum_{k=1}^{N_i} (r_i-\hat{\mu}_i)^2
\end{equation}
instead. We shall use either the biased or unbiased estimator in the following sections when appropriate.

\subsection{Approximating stylized facts using the (D$\lambda$)-model}
\label{sec:monte-carlo-simul}



\begin{figure*}[h]
\centering
\includegraphics[width=3.5in]{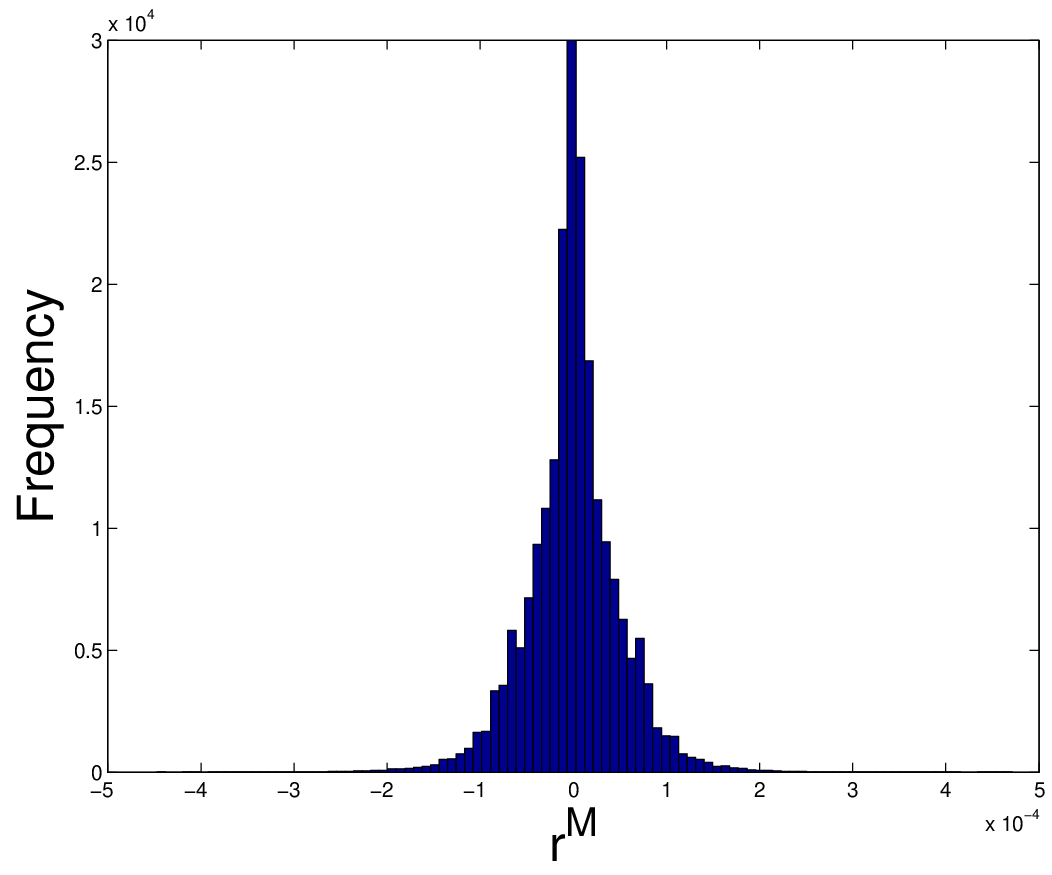}
\caption{Histogram of returns for the approximating process with $w=3$s. }
\label{Fig:HistMP3}
\end{figure*}

\begin{figure*}[h]
\centering
\includegraphics[width=3.5in]{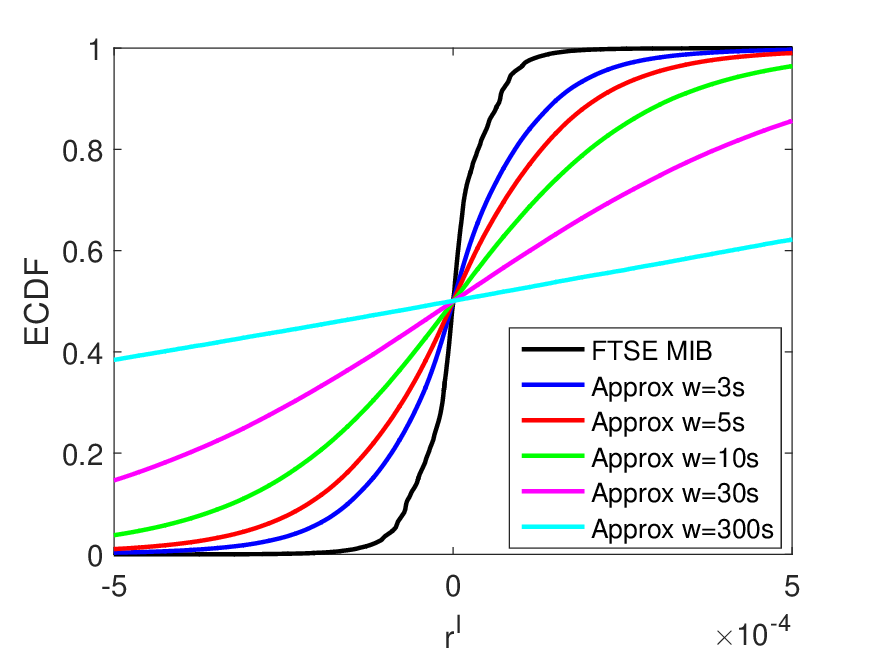}
\caption{Approximation of the empirical cumulative distribution function for FTSE MIB returns $r^I$. }
\label{Fig:Convergenza}
\end{figure*}

\begin{figure*}[h]
\centering
\includegraphics[width=4.5in]{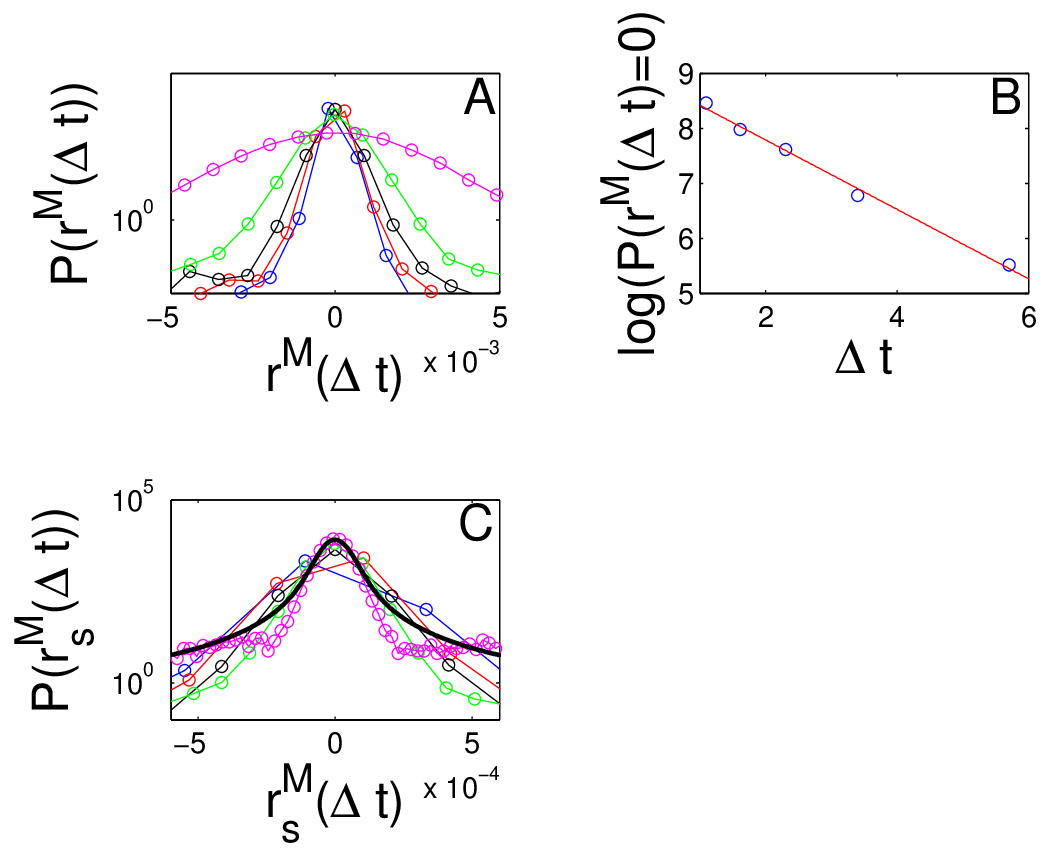}
\caption{(A) Histogram of the returns for the simulation described in the text observed at different time intervals, namely, $\Delta t=$ $3$ s (blue),  $5$ s (red), $10$ s (black), $30$ s (green) and $300$ s (purple); (B) Probability of zero returns as a function of the time sampling interval $\Delta t$, the slope of the straight line is $0.63\pm 0.01$; (C) scaled empirical probability distribution and comparison with the theoretical prediction given by Eq.\eqref{Eq:scaling2} (black solid line).}
\label{Fig:ScalingMP}
\end{figure*}

A first Monte Carlo simulation of the (D$\lambda$)-model was performed by considering a
trading day divided into a number of intervals of length $w=3,5,10,30,300$ s. The parameters $\hat{\lambda}_i$, $\hat{\mu}_i$ and $\tilde{\sigma}^2_i$ were estimated as explained above. \textcolor{black}{Note that we use the unbiased estimator $\tilde{\sigma}_i$ from (\ref{eq:1}).} In the following, we shall focus on estimates based on the FTSE MIB index. Fig \ref{Fig:HistMP3} displays the histogram of simulated returns and can be compared to Fig \ref{Fig:HistIndice}. In Fig \ref{Fig:Convergenza}, we empirically show that the simulation gives a better fit for the empirical returns of the index as $w$ becomes smaller. This is an encouraging result meaning that it will be useful to study the convergence of the approximation by means of measure-theoretical probabilistic methods.
In order to show that this approximation is able to reproduce the approximate stylized facts described above, Fig \ref{Fig:ScalingMP} shows the scaling relations discussed in section \ref{SubSecFTSEMIBIndex} for the simulation with $w=10$ s.
\textcolor{black}{The null hypothesis of normal distribution has been tested with the Kolmogorov-Smirnov, the Jarque-Bera and the Lilliefors test. The results reported in Table \ref{Fig14a} show that the null hypothesis is always rejected.}

One can see from Fig \ref{Fig:ScalingMP}(b) that an OLS index estimate $\hat{\alpha}_L=1.59$ is recovered from the simulation instead of 1.72 for the real index. The scaling given in Eqs. \eqref{Eq:scaling1}, \eqref{Eq:scaling2} is presented in Fig \ref{Fig:ScalingMP}(c), one can see that the approximate scaling still holds for the simulated data.
\textcolor{black}{The null hypothesis of identical distribution has been tested with the Kolmogorov-Smirnov test, and the results have been shown in Table \ref{Tab:Fig14c}. It is worth noting that the null hypothesis of identical distribution is always rejected but the statistical value is very near to the critical value.  }

\begin{figure*}[h]
\centering
\includegraphics[width=3.5in]{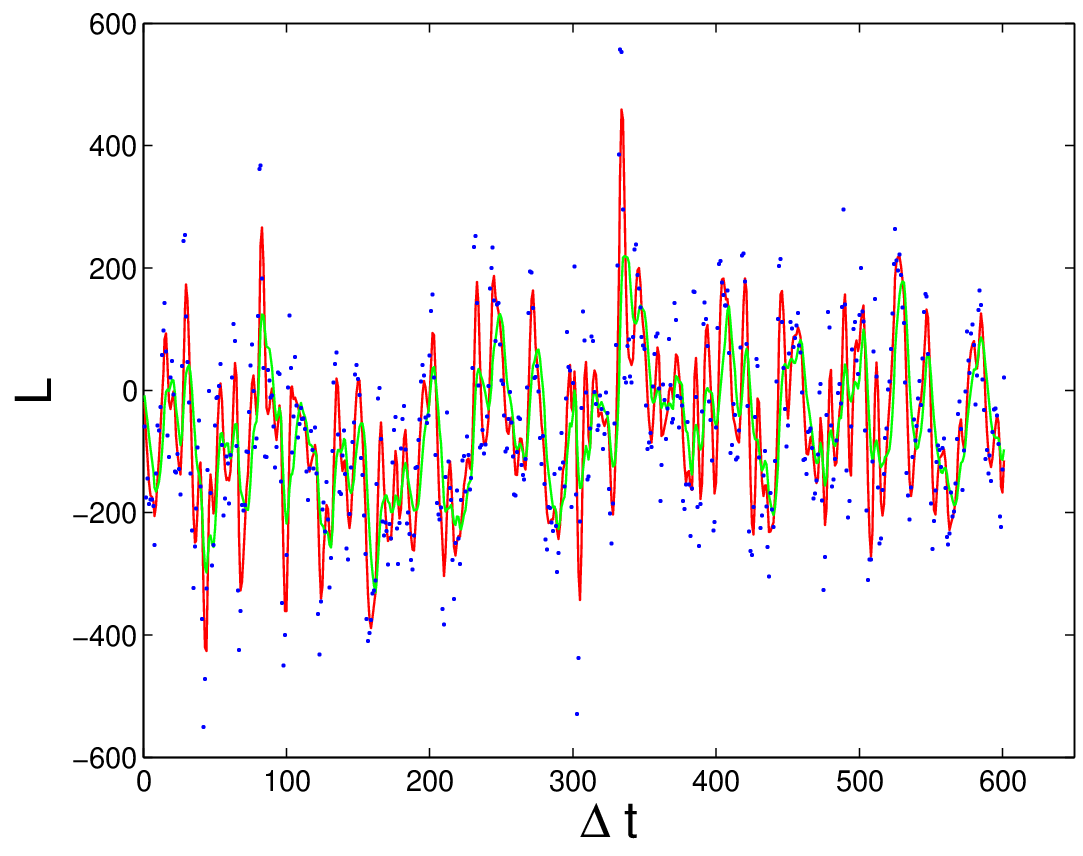}
\caption{Leverage $L$ as a function of lag $\Delta t$ for simulated data.The red and green solid lines show the leading short ($4$ lags) and lagging long ($10$ lags) square-root weighted moving average, respectively. $\Delta t$ is equal to $3$s. Also for the simulation there is no strong evidence of leverage effect.}
\label{Fig:LeverageMP3}
\end{figure*}

\begin{figure*}[h]
\centering
\includegraphics[width=4.5in]{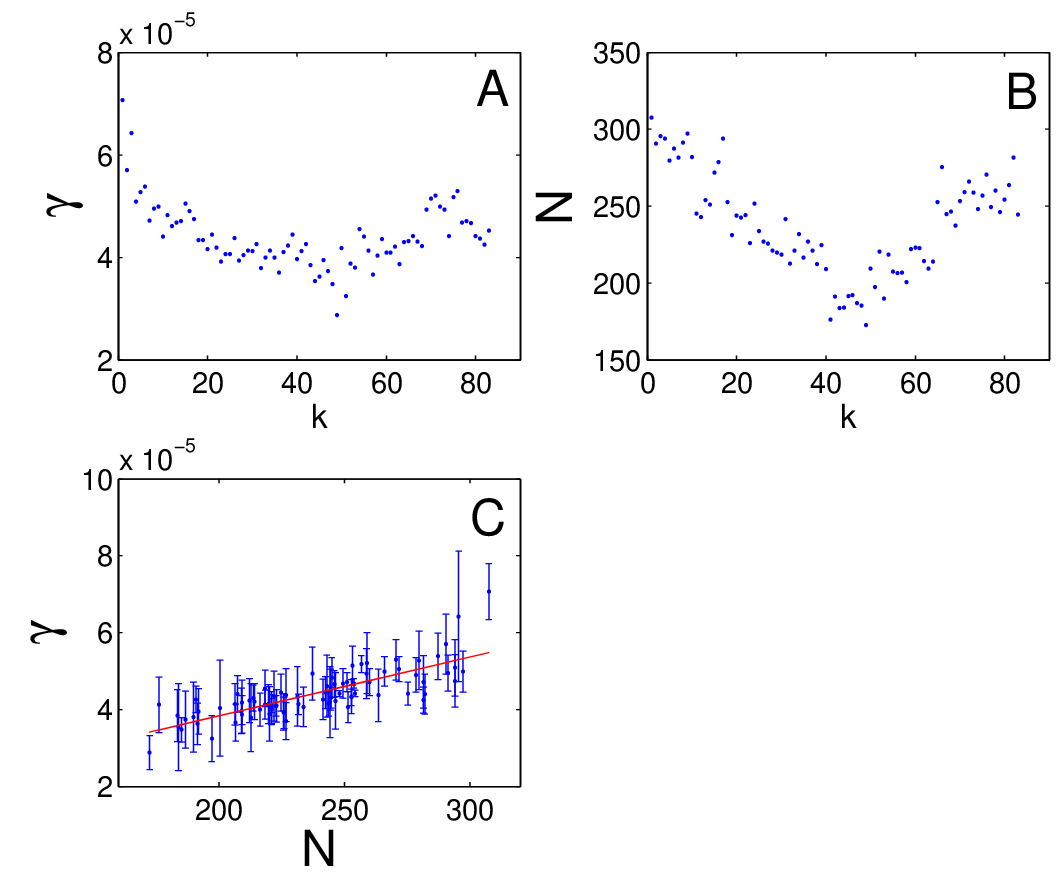}
\caption{(A) Volatility $\gamma$ as a function of $k$ for $\delta t=300$ s. (B) Activity $N$ as a function of $k$  for $\delta t=300$ s . (C) Scatter plot of volatility $\gamma$ as a function of number of trades $N$. The points are averaged over the investigated period. All the plots are for simulated data with $w=10$ s.}
\label{Fig:MP_gammaNDt300}
\end{figure*}

In Fig \ref{Fig:LeverageMP3}, we can see that there is no clear leverage effect in the simulated data as in the real case. Finally, in Fig \ref{Fig:MP_gammaNDt300}, for the simulated time series, we repeat the same analysis presented in Fig \ref{Fig:GammaNgDt300}. Given that, by construction, the non-stationary behavior of the simulated data is modeled on the non-stationary behavior of the real data, it is no surprise to find a qualitative match between the two analyses (see Figs. \ref{Fig:GammaNgDt300},\ref{Fig:MP_gammaNDt300}).

\subsubsection{The compound Poisson model with parametrized intensity (P$\lambda$)-model}
\label{sec:comp-poiss-model-1}
This model will be used for simulation later on as well as serve as a benchmark model when testing model selection criteria. As empirical results about the trading intensity suggest a daily seasonality, this model assumes that the step function in the (D$\lambda$) model is parametrized by a quadratic function:
\begin{align}
  \lambda_{a,b,c}(t) = at^2 + bt + c, \quad t \in [0,1].
\end{align}
Of course, this parametrization can be easily replaced by more complicated functions. Since $\lambda$ needs to be positive and convex, we also have the conditions
\begin{align}
  a > 0 \text{ and } c > \frac{b^2}{4a}.
\end{align}
Similar to the (D$\lambda$)-model, the log-likelihood for the (P$\lambda$)-model is given by
\begin{equation}
  \label{eq:4}
  \mathcal{L}^P_i(a,b,c, \mu_i, \sigma_i) = -\lambda_{a,b,c}(t_{i-1}) (t_{i}-t_{i-1}) + \ln(\lambda_{a,b,c}(t_{i-1})) N_i(t_i) + \sum_{k=1}^{N_i(t_i)} \ln(p_{\mu_i,\sigma_i}(R_k^{(i)})).
\end{equation}
While the maximum likelihood estimators for $\mu_i$ and $\sigma_i$ are the same as for the (D$\lambda$) case, the maximum likelihood estimators for $a,b,c$, which determine the form of $\lambda$, cannot be obtained in closed form. As a consequence, a numerical optimization method needs to be applied to estimate those parameters.

\subsubsection{The ACD model}
\label{sec:acd-model}
The autoregressive conditional duration model was first proposed by Engle and Russell \cite{Engle98}. We will consider a model for the durations between events only without marks: Let $(\varepsilon_i)_{i \in \mathbb{N}}$ be a sequence of i.i.d. random variables. The autoregressive conditional duration (ACD) model is defined as
  \begin{align}
    \label{eq:5}
    x_i &= \psi_i\varepsilon_i\\
    \psi_i &\equiv \psi_i(x_{i-1}, \ldots, x_1; \theta) := \mathbb{E}\left[ x_i \vert x_{i-1}, \ldots, x_1\right].
  \end{align}
The innovations $(\varepsilon_i)$ are assumed to follow an exponential distribution, i.e. $\varepsilon_i \sim \text{Exp}(1)$, and $\psi_i$ has the following representation
\begin{equation}
  \label{eq:9}
  \psi_i := \omega + \sum_{j=0}^m \alpha_jx_{i-j} + \sum_{j=0}^q \beta_j\psi_{i-j},
\end{equation}
where $\omega > 0$, $\alpha_i \geq 0$ and $\beta_i \geq 0$ for all $i$. We will call this model $\text{ACD}(m,q)$. For given duration data $\{x_1, \ldots, x_n\}$ the log-likelihood function is given by

\begin{equation}
  \label{eq:11}
  \mathcal{L}^{\text{ACD}}(\omega, \alpha_1, \ldots, \alpha_m, \beta_1, \ldots, \beta_q) = - \sum_{i=1}^n \left[ \ln \psi_i + \frac{x_i}{\psi_i}\right]
\end{equation}
(see p. 104 in \cite{Hautsch12}).

\subsection{MLE and goodness of fit}
\label{sec:mle-goodness-fit}
Before we turn our attention to the actual model selection procedure, it is useful to get a rough idea about how well the underlying MLE method works for the three model classes. We would also like to ensure that the MLE method works reasonably well since a poor ML fit might compromize the quality of the order selection. Due to asymptotic results, we expect that goodness of fit and correctness of the model selection procedure should improve with increasing size of the underlying sample. As these two effects are closely related, it is hard to quantify them separately.\\
In the next sections, we give a detailed explanation on the simulation procedure and on how the parameter estimation is implemented. Based on that, we run a MLE on previously generated mock data. As we know the true parameter values, we can easily calculate the mean squared error (MSE) as measure for the goodness of fit.

\subsubsection{Compound Poisson models}
\label{sec:simulation}

\paragraph{Simulation}
\label{sec:comp-poiss-models}
The simulation algorithm essentially uses the (P$\lambda$)-model. For simplicity we will choose the time interval $[t_0, T]$ to be $[0,1]$. For the simulation we set an equidistant grid $0=t_0 < t_1 < t_2 < \ldots < t_n=1$ on the time interval. Thus, the interval $[0,1]$ is divided into $n$ subintervals. For $i\in \{1, \ldots, n\}$ the parameters $\mu_i$, $\sigma_i$ and $\lambda_i$ on the subinterval $[t_{i-1}, t_i)$ are chosen to be
\begin{gather}
  \mu_i = 0, \quad \sigma_i = 1  \quad \text{ and } \quad\lambda_i = \lambda(t_{i-1}) \;\forall i\in \{1, \ldots, n\} \text{, where }\\
  \lambda(t) := 4(\lambda_{\text{max}} - \lambda_{\text{min}})(t-0.5)^2 + \lambda_{\text{min}}, \; \forall t \in [0,1] \text{ and } \lambda_{\text{min}}, \lambda_{\text{max}} > 0 \text{ constant}. \label{eq:3}
\end{gather}
The functional form of $\lambda$ is inspired by the empirical findings in the previous sections and should account for the observed seasonality in a simple way. Of course, the functional form of $\lambda$ can be easily replaced by more complex functions. We have chosen $\lambda_{\text{min}}=10$0 and $\lambda_{\text{max}}=10000$. Note that the $\{\lambda_i\}$ form a step function approximation of the parabola in (\ref{eq:3}). For different grid sizes, we simulate with sample size $1000$ each.

\paragraph{Fitting}
\label{sec:fitting-1}
The fitting will be carried out using different grid sizes. Note that the grid size to be used in fitting is bounded from above by the length of the entire time interval (in our case 1). However, as we would like to emulate the behavior of the intensity which was observed in empirical data, i.e. high intensity at the beginning and at the end of the trading day and relatively low intensity in the middle of the day. Consequently, we need at least 3 subintervals to have a piecewise constant function that fulfils these conditions on the time interval. Further, the smallest eligible grid size is bounded from below by the maximal distance between neighbouring data points within the data set. Otherwise, there are subintervals which do not contain any data points. In such cases, the estimation formulas in (\ref{eq:2}) would fail.\\
More precisely, for the maximal distance $\Delta_{\text{max}}$ between two consecutive data points within a given sample, the finest valid equidistant grid has at most $\left\lfloor \frac{1}{\Delta_{\text{max}}}\right\rfloor$ subintervals. Therefore, we will consider a list of candidate models on grids which correspond to $n=3, 4, \ldots, \left\lfloor \frac{1}{\Delta_{\text{max}}}\right\rfloor$ subintervals on the interval $[0,1]$.\\

For the (D$\lambda$) model, the estimators are given in closed form in (\ref{eq:2}) and the likelihood value is easily calculated via Eq (\ref{eq:12}) and subsequently used for the calculation of the IC. We decide to use the unbiased estimator $\hat{\sigma}^2_i$: Since we are mainly interested in model selection, we would like to ensure that we work with the optimal value of the log-likelihood when calculating the IC (see also \ref{sec:model-selection}).\\
In order to fit the (P$\lambda$) model, we assume that the estimates for $\{\mu_i\}, \{\sigma_i\}$ and $\{\lambda_i\}$ for the (D$\lambda$)-algorithm are already calculated and can be used as an input for the estimation of the (P$\lambda$)-model. As mentioned previously, the estimators for $\mu_i$ and $\sigma_i$ coincide in both models and no further calculation is needed for these parameters. It remains to solve the following minimization problem:
\begin{gather}
  (\hat{a}, \hat{b}, \hat{c}) = \argmin_{a,b,c \in \mathbb{R}} \left[-\sum_{i=1}^n\mathcal{L}^P_i(a,b,c, \mu_i, \sigma_i) \right] \quad \text{s.t.} \quad  a > 0 \text{ and } c > \frac{b^2}{4a}
\end{gather}
A reasonable choice of the starting value for the minimization algorithm can be easily obtained by the least-squares fit of the parabola to the $\{\lambda_i\}$ values of the (D$\lambda$) case, which already gives a fairly good approximation of the parabola. In case the initial values obtained by this method do not lie in the admissible set, a change of signs for $a$ or a shift of the parabola may be applied.\\
Note that the estimation of the (P$\lambda$)-model requires a grid with at least 4 grid points, i.e. 3 subintervals on which $\lambda_1, \lambda_2, \lambda_3$ are estimated using the (D$\lambda$)-model. This ensures that the parabola is well determined. However, as mentioned before, this condition is not restrictive and covers all models on which we would like to run model selection.
\subsubsection{ACD model}
\label{sec:acd-model-1}
For both simulation and MLE of ACD models we use the \texttt{R} package \texttt{ACDm} written by Markus Belfrage \cite{acdm}. The model selection analysis for the ACD model follows the Monte Carlo experiment conducted in \cite{Javed_Mantalos_2013}. We consider model orders $m,q \in \{1,2\}$ and Table \ref{acdParam} shows the choice of parameters for the simulation.

\subsubsection{Numerical results}
\label{sec:numemrical-results}
We use the MSE as a measure for the goodness of fit: Let $\theta$ be a generic model parameter to be estimated and $\hat{\theta}$ the corresponding estimator. Given $N=1000$ samples and $\hat{\theta}^{(k)}$, $k=1,\ldots,N$, the estimates for each sample we calculate the mean squared error to be
\begin{equation}
  \label{eq:13}
  \text{MSE($\theta$)} = \mathbb{E}\left[\vert\theta - \hat{\theta}\vert^2\right] = \frac{1}{N} \sum_{k=1}^N \vert\theta - \hat{\theta}^{(k)}\vert^2.
\end{equation}
\paragraph{Compound Poisson models}
\label{sec:comp-poiss-models-1}
We have to point out first that the distance in Eq (\ref{eq:13}) has to be understood as a functional distance. To be more precise, we choose the $L^2$-distance between the true step function intensity and the estimated one:
\begin{equation}
  \label{eq:14}
  \mathbb{E}\left[\vert\theta - \hat{\theta}\vert^2\right] = \mathbb{E}\left[\Vert\theta - \hat{\theta}\Vert_{L^2}^2\right]
\end{equation}
The cases of $\mu$ and $\sigma^2$ are the easier ones, as we just need to calculate the distance between a step function and a constant: For the step functions with values $\{\mu_i\}$ on the fitting grid  $t_1 < t_2 < \ldots < t_n$ Eq (\ref{eq:14}) can be further written as
\begin{align}
  \label{eq:15}
  \mathbb{E}\left[\Vert\mu - \hat{\mu}\Vert_{L^2}^2\right] &= \frac{1}{N} \sum_{k=1}^N \Vert\mu - \hat{\mu}^{(k)}\Vert_{L^2}^2 = \frac{1}{N} \sum_{k=1}^N \int_0^T (\mu(t) - \hat{\mu}^{(k)}(t))^2 \mathrm{d} t\\
&=\frac{1}{N} \sum_{k=1}^N \sum_{i=2}^n (\mu - \hat{\mu}^{(k)}_i)^2(t_i-t_{i-1})\\
\end{align}
and in the same way for $\sigma^2$.\\
Concerning the intensity function, we have to merge the simulation grid $t_1^s < t_2^s < \ldots < t_m^s$ with the fitting grid $t_1^f < t_2^f < \ldots < t_r^f$. After reordering and relabeling, we can calculate the MSE on the merged grid $t_1 < t_2 < \ldots < t_n$ via
\begin{equation}
  \label{eq:16}
  \mathbb{E}\left[\Vert\lambda - \hat{\lambda}\Vert_{L^2}^2\right] = \frac{1}{N} \sum_{k=1}^N \sum_{i=2}^n (\lambda_i - \hat{\lambda}^{(k)}_i)^2(t_i-t_{i-1}).
\end{equation}
The numerical results we present here are exemplarily for $N=1000$ samples of data simulated from a grid containing $30$ subintervals.\\

\begin{figure}[h]
  \centering
  \includegraphics[scale=0.8]{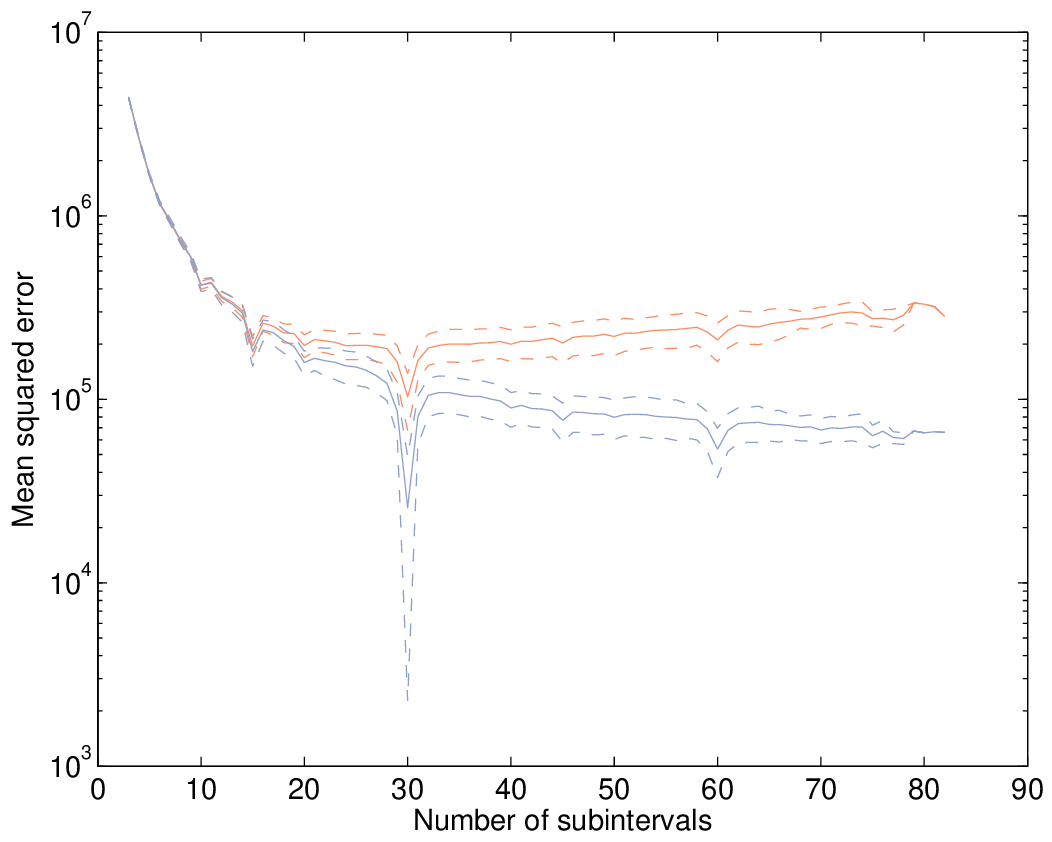}
  \caption{Plot of the mean squared error (MSE) of the estimation of the intensity function for the (D$\lambda$) model (orange lines) and for the (P$\lambda$) model (blue lines) respectively. The graph shows the MSE together with dashed lines indicating the size of the first standard deviation from the mean as a function of the underlying number of intervals of the fitting grid. The true model for the simulation originally used 30 subintervals. The MSE is calculated as a squared $L^2$ distance between the estimated and the true intensity function (see also Eq. (\ref{eq:16})).}
  \label{fig:mseLambda}
\end{figure}

Table \ref{mseMuSIgma} shows summary statistics of $\mu$ and $\sigma^2$, where the summary statistics were calculated over the set of fitting grids. The MSE for the $\mu$ and $\sigma^2$ are comparably small.\\
For the intensity function $\lambda$ we plot the MSE against the number of subintervals used for fitting in Fig \ref{fig:mseLambda}. Starting from a small number of subintervals, the MSE decreases sharply before it reaches its optimum at 30, the true number of subintervals from the simulation. Number of subintervals above 30 give a larger MSE and, in the case of the (D$\lambda$) model, instabilities of over parametrization even lead to an increasing MSE.\\
Concerning goodness of fit, we can see that the MSE of the (P$\lambda$) model is consistently smaller than the MSE of the (D$\lambda$) model. This is to be expected as by construction of the experiment the (P$\lambda$) model is the true model and gives a better fit to the data.\\
Moreover, we can observe that apart from the optimum at 30 there are ``preferred'' numbers of subintervals at 10, 20, 45, 60. This is crucial for the explanation of the behavior of model selection as the relationship between goodness of fit and number of subintervals in the region below the optimal number is not monotone.\\

One might be concerned about the large values of the MSE of the $\lambda$ estimation. However, first note that the sample size is neither controlled by the choice of simulation grid size nor the fitting grid size. The sample size is determined by the value of the intensity $\lambda$. Consequently, if the fitting grid is already sufficiently fine, the sample size is approximately of the same order. Since, the sample size does not change much for finer fitting grids, we therefore cannot expect to observe any convergence of the MSE to $0$ in Fig \ref{fig:mseLambda}.
\\
Second and more importantly, the size of the MSE can be estimated by the expected fluctuations of the estimator $\hat{\lambda}$. The MSE can be estimated from below by the ideal situation when the simulation and fitting grid are identical. Without loss of generality, we assume an equidistant simulation grid with grid size $w=t_i-t_{i-1}$ and rewrite Eq (\ref{eq:16}):
\begin{equation}
  \label{eq:17}
  \mathbb{E}\left[\Vert\lambda - \hat{\lambda}\Vert_{L^2}^2\right] \geq w\sum_{i=2}^n \mathbb{E}\left[(\lambda_i - \hat{\lambda}_i)^2\right] = w\sum_{i=2}^n \text{Var}\left[\hat{\lambda}_i\right] =\frac{1}{w} \sum_{i=2}^n \text{Var}\left[N_i\right],
\end{equation}
where we have used the definition of the estimator in (\ref{eq:2}) and that the number of events in an interval of size $w$ is Poisson distributed: $N_i \sim \text{Poi}(\lambda w)$. We finally get that
\begin{equation}
  \label{eq:18}
  \mathbb{E}\left[\Vert\lambda - \hat{\lambda}\Vert_{L^2}^2\right] \geq \frac{1}{w} \sum_{i=2}^n \text{Var}\left[N_i\right] = \frac{1}{w} \sum_{i=2}^n \lambda_i w \approx \frac{1}{w} \int_0^1 \lambda(t) \; \mathrm{d} t,
\end{equation}
where we approximate the integral of the step function by the integral of the smooth intensity parametrization in Eq (\ref{eq:3}). For our numerical example we have $\frac{1}{w} = 30$ and $\lambda_{\text{min}}=100$ and $\lambda_{\text{max}}=10000$. An explicit calculation of above integral gives the rough estimate
\begin{equation}
  \label{eq:19}
  \mathbb{E}\left[\Vert\lambda - \hat{\lambda}\Vert_{L^2}^2\right] \gtrsim 30\cdot 3400 = \mathcal{O}(10^5),
\end{equation}
which is of about the same order of magnitude observable in Fig \ref{fig:mseLambda}.

\paragraph{ACD model}
\label{sec:acd-model-2}
In the ACD case we have a simple parameter vector $(\omega, \alpha_1, \ldots, \alpha_m, \beta_1, \ldots, \beta_q) \in \mathbb{R}^{1+m+q}$, Therefore, we can use the formula given in Eq (\ref{eq:13}) for each scalar valued parameter. The results can be seen in Table \ref{mseACD}. The largest sample size ensures that the MSE are comparably low for each model. The largest contribution to the MSE comes from the $\omega$ parameter. An even closer look shows that the MSE of the $\beta$ parameter(s) is of different order depending on the model order $q$. In the case $q=1$, the MSE of the $\beta$ parameter is of the same size as the $\alpha$ parameter(s). However, in the case of $q=2$, the order of the MSE of the $\beta$ parameters are significantly larger than the MSE of the $\alpha$ parameters (by a factor of $10$ in the $\text{ACD}(1,2)$ case and by a factor of $100$ in the $\text{ACD}(2,2)$ case).

\subsection{Information criteria and model selection}
\label{sec:model-selection}
Starting off from the estimation results in the previous section, we would like to analyse how effective model selection based on information criteria (IC) performs for both the coumpounds Poisson models and the ACD model.\\
As seen in the previous Monte Carlo simulation choosing smaller values of $w$, i.e. increasing the number of model parameters, gives better fits and the model is able to capture all distributional properties of the quantity of interest. However, a model containing a large number of parameters is likely to be over fitted. A quantitative method to resolve this trade-off situation is to apply IC.
In the following, we will consider three of the most common information criteria:\\
  For a given model fitted to data via MLE let $\mathcal{L}$ be the maximal log-likelihood value, $k$ the number of parameters and $T$ be the sample size of the data set. Then we define:
  \begin{enumerate}
  \item \textbf{Akaike's information criterion (AIC)} (see \cite{Akaike73})
    \begin{equation}
      \label{eq:6}
      \text{AIC} = -2\mathcal{L} + 2k
    \end{equation}
  \item \textbf{Bayesian information criterion (BIC)} (see \cite{Schwarz78})
    \begin{equation}
      \label{eq:7}
      \text{BIC} = -2\mathcal{L} + k\ln(T)
    \end{equation}
  \item \textbf{Hannan and Quinn information criterion (HQ)} (see \cite{Hannan79} and \cite{Hannan80})
    \begin{equation}
      \label{eq:8}
      \text{HQ} = -2\mathcal{L} + 2k \ln(\ln(T))
    \end{equation}
  \end{enumerate}
Note that the information criteria under consideration penalize the log-likelihood value for increasing number of parameters $k$. Among several candidate models, one chooses the model with the smallest IC value. A time grid $t_0 < t_1 < \ldots < t_n$ is given and divides the overall time interval in $n$ subintervals. Recall that we from Section \ref{sec:fitting-1} that we do not consider $n\in \{1,2\}$. Then the (D$\lambda$)-model has in total $k=3n$ parameters with $n\in \{3, 4, \ldots\}$. This will also be the true number of parameters we expect the IC to choose. In the same way we have for the (P$\lambda$)-model $k=2n+3$ parameters with $n\in \{3, 4, \ldots\}$.

\subsubsection{Numerical results}
\label{sec:numerical-results-1}

\begin{figure*}[h]
  \centering
  \includegraphics[scale=0.8]{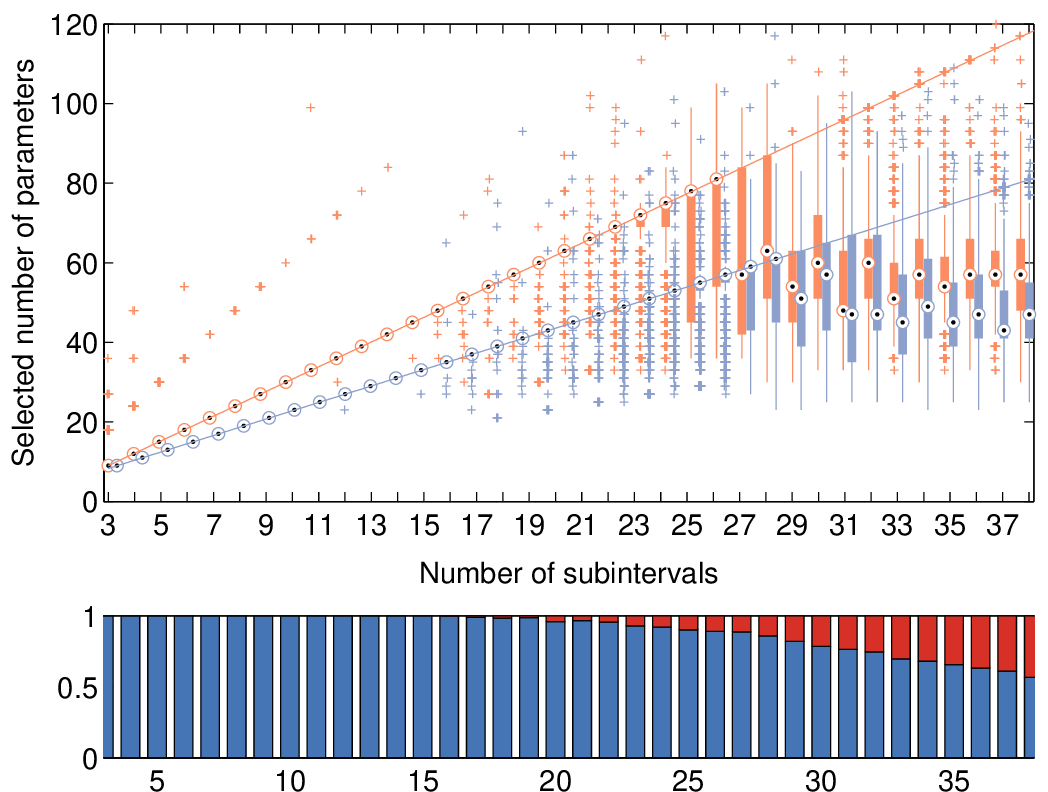}
  \caption{
The lower plot shows the ratio of samples which allow the true model to be among the set of models from which the IC may choose from, in other words there is no misspecification (blue areas). This ratio decreases and for finer discretization there are more cases of model selection under misspecification (red areas). The sum of blue and red areas is 100\%. \\
The upper plot shows that the model selection using the AIC for the (D$\lambda$)-model (orange box plot) closely follows the reference line indicating $3n$ ($n=$ number of subintervals) for small $n$, before deviating for larger $n$. The same holds for the (P$\lambda$)-model (blue box plot) and its corresponding reference line $2n+1$. The number of subintervals for which both box plots deviate from their respective reference lines is around $n=25$ to $n=27$. In the region $n < 15$, there are several outliers which are almost all overestimates.
}
  \label{fig:AIC_boxplot}
\end{figure*}

\begin{figure*}[h]
  \centering
  \includegraphics[scale=0.8]{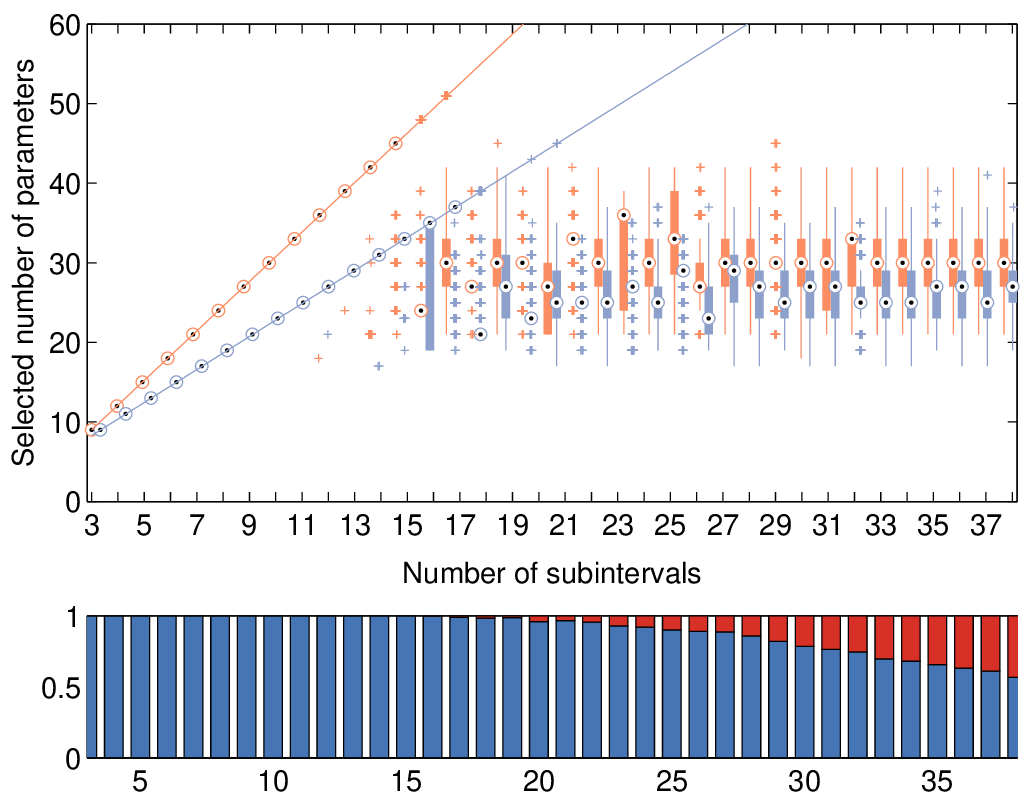}
  \caption{The lower plot shows the ratio of samples which allow the true model to be among the set of models from which the IC may choose from, in other words there is no misspecification (blue areas). This ratio decreases and for finer discretization there are more cases of model selection under misspecification (red areas). The sum of blue and red areas is 100\%. \\
The upper plot shows that the model selection using the BIC for the (D$\lambda$)-model (orange box plot) closely follows the reference line indicating $3n$ ($n=$ number of subintervals) for small $n$ before deviating for larger $n$. The same holds for the (P$\lambda$)-model (blue box plots) and its corresponding reference line $2n+1$. The number of subintervals for which both box plots deviate from their respective reference lines is around $n=15$ to $n=17$.
}
  \label{fig:BIC_boxplot}
\end{figure*}

\begin{figure*}[h]
  \centering
  \includegraphics[scale=0.8]{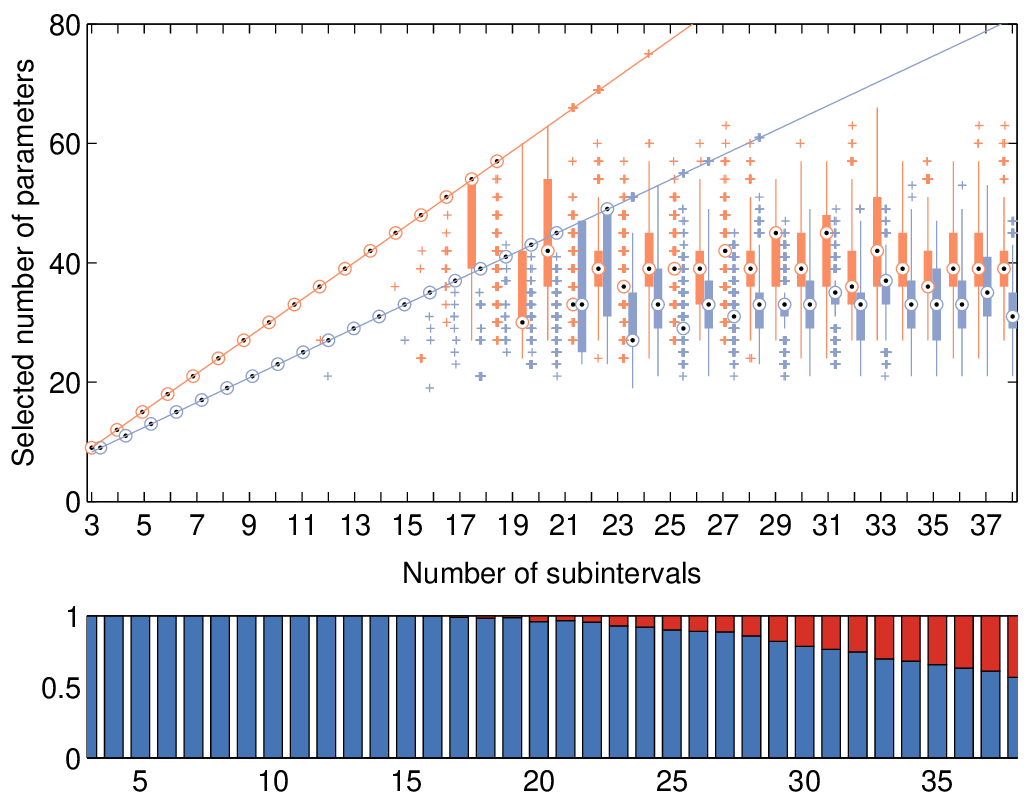}
  \caption{The lower plot shows the ratio of samples which allow the true model to be among the set of models from which the IC may choose from, in other words there is no misspecification (blue areas). This ratio decreases and for finer discretization there are more cases of model selection under misspecification (red areas). The sum of blue and red areas is 100\%. \\
The upper plot shows that the model selection using the HQ for the (D$\lambda$)-model (orange box plot) closely follows the reference line indicating $3n$ ($n=$ number of subintervals) for small $n$ before deviating for larger $n$. The same holds for the (P$\lambda$)-model (blue box plots) and its corresponding reference line $2n+1$. The number of subintervals for which both box plots deviate from their respective reference lines is around $n=18$ to $n=20$.
}
\label{fig:HQ_boxplot}
\end{figure*}

\paragraph{Compound Poisson models}
\label{sec:d}
Figs \ref{fig:AIC_boxplot}, \ref{fig:BIC_boxplot} and \ref{fig:HQ_boxplot} show box plots of the model selection results of the AIC, BIC and HQ respectively. In each box plot, the orange and blue box plot correspond to the results of the (D$\lambda$)- and (P$\lambda$)-model respectively. The horizontal axis shows the number of subintervals used in the simulation grid. On the vertical axis are the selected number of parameters after the parameter estimation of the (D$\lambda$)- and (P$\lambda$)-models using different discretizations of $[0,1]$. A single box in the box plots extends from the 25th percentile to the 75th percentile and the dot  indicates the median. The whiskers have a maximum length of 1.5 times the box length and extend to the outermost point which is not considered as outlier. The crosses indicate outliers.\\
Below the box plots, bars indicate the ratio of samples which allow model selection under correct specification (blue) and under misspecification (red): In our setting, we speak of model selection under misspecification if the correct model is not contained in the set of selectable models and cannot be chosen by the IC. If this is not the case, i.e. the correct model can potentially be chosen by the IC, we call it model selection under correct specification.\\

The results for the (D$\lambda$) and (P$\lambda$) model are very similar. Common for all three IC is that for small parameter numbers below 15 the model selection works well: the distributions of the selected orders are concentrated and closely follow the $3n$ or $2n+3$ reference line respectively, where $n$ is the number of subintervals. For very large parameter numbers one can observe that the selected model orders remain distributed around a maximum model order and stop to follow the linear trend of the reference line. This is rather due to the limitations of our MC setup than the inherent property of the IC: As described in Section \ref{sec:comp-poiss-models}, we only work with equidistant grids when applying the model selection procedure. The finest grid which can be used for fitting is determined by the maximal distance $\Delta_{\text{max}}$ between two consecutive points within a sample. On the other hand, $\Delta_{\text{max}}$ is related to the minimal value of $\lambda$ in the middle of the interval., depending on how small we choose the simulation grid size $\Delta_{\text{sim}}$. This means that whenever $\Delta_{\text{max}} > \Delta_{\text{sim}}$, the true model is not contained in the pool of models from which the IC may choose from. In other words, we have a case of model selection under misspecification. The bar plots show that first cases occur at around $n=20$ and go up to a ratio of about $50\%$ for the finest grid in the analysis.\\
Another look at Fig \ref{fig:mseLambda} hints that the general rule ``the more parameters, the better the fit'' is not  entirely true: we can observe that the relation between grid size and MSE is not entirely monotone. This is due to the fact that the fit of the specific model does not only depend on the number of parameters, but also to some extent on the position of the grid. As a consequence, under misspecification, the selected order does not necessarily correspond to the finest available grid size above $\Delta_{\text{sim}}$. This might explain the ``plateaus'' on the model selection results for large parameters.\\

Between the region of very small and very large parameters the IC exhibit quite different behaviors according to their intrinsic tendency of under- and overfitting, which will be described in the following:\\
The AIC tends to overestimate the number of parameters. It allows outliers (in the region of $n \leq 22$) as well as a larger number of cases of the model selection to lie above the reference line (in the region of $n\geq 23$). In contrast, the selected model orders of the BIC and HQ are either on the reference line or strictly below the reference line. In other words BIC and HQ tend to underestimate. Additionally,  we can see that for the AIC the boxplot starts to deviate from the reference line starting around $n=25$ to $n=27$ and the BIC and HQ deviate earlier around $n=15$ and $n=20$ respectively. Especially, for $n < 27$ the underestimation in the BIC and HQ case is not attributable to the behaviour of model selection under misspecification, as the ratio of model selection under misspecification is rather low. Based on our results, If the IC were to be ordered by their parsimonious character, the BIC would be the more parsimonious whereas the AIC the least. \\
The above observations show that the model selection using any of the three IC  works quite well as long as the true model is actually retrievable. The AIC tends to overestimate, but the model selection results are closest to the reference line of true parameters compared to the other two IC.


\paragraph{ACD model}
The results of the model selection experiment can be found in Tables \ref{ACD11} to \ref{ACD22}. The numbers are success rates in percent of the respective IC to select the correct model from which the simulation data was generated from. The qualitative behaviour of the IC are not surprisingly similar to the findings for the GARCH model in \cite{Javed_Mantalos_2013}.\\

A closer look at Table \ref{ACD11} shows that the success rate of the IC is exceptionally good in the case of $\text{ACD}(1,1)$ data. Even for a small sample size all information criteria are able to detect the correct model order in the majority of cases. The tendency to under fit works in favour for the BIC and to some extent also for the HQ. For the same reason, the success rates for the AIC are relatively low due to its overfitting property.\\
A similar behaviour can be observed for $\text{ACD}(2,1)$ in Table \ref{ACD21}: Although the IC underestimate the model for smaller sample sizes as a $\text{ACD}(1,1)$ model, they improve for large sample sizes.\\
In both the $\text{ACD}(1,1)$ and the $\text{ACD}(2,1)$ case, i.e. the cases for $q=1$, the behaviour of the model selection is acceptable: a reasonably large sample size, which is of the order of a typical intra day trading data sample, ensures a sufficiently large success rate in detecting the correct model. Unfortunately, this cannot be said about the case $q=2$:\\
In the first example of $\text{ACD}(1,2)$ data in Table \ref{ACD12}, we see that the correct model order is never detected in the majority of cases even for large sample sizes. The best success rates are the ones of the AIC again due to its overfitting tendency. This may be concerning, as this shows that despite the fact that $\text{ACD}(1,2)$ and $\text{ACD}(2,1)$ have the same number of parameters the model selection behaviour is far from comparable.\\
In comparison, the results for the $\text{ACD}(2,2)$, the most complex model in our experiment, are even more critical: Not only are the IC unable to detect the correct model in most of the cases even with large samples, but the best success rates, again from the AIC, are below 20\%.\\

As mentioned in Section \ref{sec:acd-model-2}, the cases where model selection fails align with relatively high MSE of the $\beta$ parameters for $q=2$: The contribution of the MSE of the $\omega$ parameter is not as important, as this parameter is included in all models. However, the increase in MSE when moving from $q=1$ to $q=2$ might be one of the factors explaining the discrepancy in model selection between $q=1$ and $q=2$. This part of our MC experiment suggests that parameters which are harder to estimate compared to other model parameters (in our case $\alpha$ vs. $\beta$ parameters or in other words moving average vs. autoregressive parameters in Eq (\ref{eq:9}), might also be less likely to be detected by model selection.

\section{Conclusions}
\label{SecConclusions}
In this paper, we addressed two questions. The first one concerns to so-called stylized facts for high-frequency financial data. In particular, do the statistical regularities detected in the past still hold? We cannot give a negative answer to this question. Indeed, we find that some of the scaling properties for financial returns are still approximately
satisfied. Most of the studies we refer to concerned a different market (the US NYSE) and were performed several years ago. However, one of the first econophysics papers (if not the first one) concerned returns in the Italian stock exchange (see \cite{Mantegna91})  and, for this reason, we decided to focus on this market.

The second question is: Is it possible to approximate the non-stationary behavior of intra-day tick-by-tick returns by means of a simple phenomenological stochastic process? We cannot give a negative answer to this question, so far. In Section \ref{SecModel}, we present a simple non-homogeneous normal compound Poisson process and we argue that it can approximate empirical data. The cost for simplicity is potential over-fitting as we have to estimate many parameters, but the outcome is a rather accurate representation of the real process. Whether it is possible to rigorously prove convergence of the method outlined in Section \ref{SecModel} is subject to further research and it is outside the scope of the present paper. It is well-known that L\'evy processes, namely stochastic processes with stationary and independent increments, can be approximated by compound Poisson processes. The method described in Section \ref{SecModel} can provide a clue for a generalization of such a result to processes with non-stationary and non-independent increments.

Concerning the issue of overfitting, the second part of Section \ref{SecModel} shows that IC are able to detect model orders correctly to some extent when applied to simulated data. It remains to check how well the model selection method performs on empirical data. As a consequence from the numerical results, due to the high variability of model selection in the region of larger numbers of parameters it is not advisable to rely only on the IC based \textcolor{black}{model selection}. It is recommended to combine these with further cross-validation techniques.
\textcolor{black}{A similar conclusion holds for the ACD model, as model selection using IC is adversely affected by differing MLE quality for different model orders.}

\section*{Acknowledgments}
This work was partially supported by MIUR PRIN 2009 grant {\em The growth of firms and countries: distributional properties and economic determinants - Finitary and non-finitary probabilistic models in economics 2009H8WPX5$\_$002} and by an SDF fund provided by the University of Sussex.

%
%
%
%


\bibliography{MisturaBibliografia}

\newpage

%
%
%
%
%
%
%
%
%

\section*{Figure captions}


\clearpage


\FloatBarrier

\newpage

\section*{Tables}

\begin{table*}[tb]\footnotesize
	\centering
	\caption{Symbols and number of observations for the 40 assets composing the FTSE MIB index in February-March 2011}
	\label{Tab:40AzioniSymbol}
\begin{tabular}{|l|l|c|}
	\hline
  Asset           &  Symbol  &  Number of observations    \\
  \hline
A2A                &  A2A    &  17987 \\
Ansaldo STS        &  STS    &  14252 \\
Atlantia           &  ATL    &  25811 \\
Autogrill Spa      &  AGL    &  15834 \\
Azimut             &  AZM    &  14779 \\
Banco Popolare     &  BP     &  70373 \\
Bca MPS            &  BMPS   &  38005 \\
Bca Pop Milano     &  PMI    &  32132 \\
Bulgari            &  BUL    &  20164 \\
Buzzi Unicem       &  BZU    &  25236 \\
Campari            &  CPR    &  14789 \\
Diasorin           &  DIA    &  16386 \\
Enel               &  ENEL   &  73223 \\
Enel Green Power   &  EGPW   &  29305 \\
ENI                &  ENI    &  77280 \\
Exor               &  EXO    &  26108 \\
Fiat               &  F      &  84641 \\
Fiat Industrial    &  FI     &  52212 \\
Finmeccanica       &  FNC    &  31566 \\
Fondiaria-SAI      &  FSA    &  21169 \\
Generali Ass       &  G      &  60561 \\
Impregilo          &  IPG    &  16414 \\
Intesa Sanpaolo    &  ISP    &  84525 \\
Lottomatica        &  LTO    &  14313 \\
Luxottica Group    &  LUX    &  25717 \\
Mediaset           &  MS     &  32019 \\
Mediobanca         &  MB     &  37848 \\
Mediolanum         &  MED    &  17185 \\
Parmalat           &  PLT    &  30861 \\
Pirelli \& C       &  PC     &  27023 \\
Prysmian           &  PRY    &  32806 \\
Saipem             &  SPM    &  57592 \\
Snam Rete Gas      &  SRG    &  25324 \\
STMicroelectronics &  STM    &  54515 \\
Telecom Italia     &  TIT    &  49576 \\
Tenaris            &  TEN    &  36410 \\
Terna              &  TRN    &  21836 \\
Tod's              &  TOD    &  14811 \\
Ubi Banca          &  UBI    &  31541 \\
UniCredit          &  UCG    &  168433 \\
\hline
Index              & FTSE MIB & 405560\\
\hline
\end{tabular}
\end{table*}

\input{table1newtest}

\begin{table*}[tb]\footnotesize
	\centering
	\caption{Descriptive statistics for the volumes $v^h$}
	\label{Tab:StatisticaVolumi}
\begin{tabular}{|l|r|r|r|r|}
	\hline
  Assets  & mean $\times10^{4}$ & variance $\times10^{8}$ & skewness & kurtosis $\times10^{2}$\\
  \hline
A2A   &                 1.11  &  $          5.72   $              &        11.17 &             2.75    \\
STS   &                 0.11  &  $          0.05     $            &        10.86 &             2.79    \\
ATL   &                 0.16  &  $          0.09     $            &         8.79 &             2.16    \\
AGL   &                 0.15  &  $          0.09     $            &         7.97 &             1.26    \\
AZM   &                 0.13  &  $          0.05     $            &         6.10 &             0.70    \\
BP    &                 1.17  &  $          6.21   $              &        20.98 &             12.14   \\
BMPS  &                 1.69  &  $          10.05  $              &         6.98 &             1.01    \\
PMI   &                 0.52  &  $          0.67    $             &         5.64 &             0.74    \\
BUL   &                 0.53  &  $          7.33   $              &        26.99 &             13.21   \\
BZU   &                 0.16  &  $          0.07     $            &         7.05 &             1.21    \\
CPR   &                 0.18  &  $          0.08     $            &         5.66 &             0.61    \\
DIA   &                 0.03  &  $          0.28 \times10^{-2} $  &         6.33 &             0.73    \\
ENEL  &                 1.09  &  $          7.06   $              &        15.92 &             6.97    \\
EGPW  &                 0.80  &  $          2.78   $              &        12.88 &             3.60    \\
ENI   &                 0.48  &  $          2.20   $              &        78.73 &             118.40  \\
EXO   &                 0.07  &  $          0.01     $            &         5.10 &             0.49    \\
F     &                 0.62  &  $          1.68   $              &         9.31 &             2.05    \\
FI    &                 0.31  &  $          0.37    $             &         8.04 &             1.36    \\
FNC   &                 0.18  &  $          0.14    $             &        10.76 &             3.01    \\
FSA   &                 0.22  &  $          0.14    $             &        10.39 &             3.42    \\
G     &                 0.31  &  $          0.35    $             &         9.09 &             2.32    \\
IPG   &                 0.56  &  $          1.40   $              &        13.44 &             3.88    \\
ISP   &                 3.39  &  $          45.25  $              &         7.56 &             1.72    \\
LTO   &                 0.14  &  $          0.07     $            &         6.67 &             0.81    \\
LUX   &                 0.08  &  $          0.02     $            &        10.30 &             2.83    \\
MS    &                 0.41  &  $          0.54    $             &         7.23 &             1.19    \\
MB    &                 0.28  &  $          0.26    $             &         8.41 &             1.66    \\
MED   &                 0.31  &  $          0.33    $             &        10.09 &             2.29    \\
PLT   &                 1.01  &  $          8.72   $              &        31.52 &             17.87   \\
PC    &                 0.33  &  $          0.37    $             &         9.07 &             2.07    \\
PRY   &                 0.14  &  $          0.07     $            &         7.80 &             1.32    \\
SPM   &                 0.09  &  $          0.03     $            &        13.07 &             5.58    \\
SRG   &                 0.56  &  $          6.92   $              &       117.04 &             166.34  \\
STM   &                 0.29  &  $          0.32    $             &         7.21 &             1.29    \\
TIT   &                 3.26  &  $          66.70  $              &        18.81 &             11.30   \\
TEN   &                 0.17  &  $          0.09     $            &         9.18 &             2.11    \\
TRN   &                 0.83  &  $          5.89   $              &        61.35 &             64.82   \\
TOD   &                 0.02  &  $          0.08 \times10^{-2} $  &         7.52 &             1.07    \\
UBI   &                 0.23  &  $          0.17    $             &         6.87 &             1.01    \\
UCG   &                 5.63  &  $          124.95 $              &         7.82 &             1.78    \\
\hline
\end{tabular}
\end{table*}

\begin{table*}[tb]\footnotesize
	\centering
	\caption{Descriptive statistics for the trade-by-trade log-returns $r^h$. (*) On March 7$^{\mathrm{th}}$, 2011, the French firm
	LVMH launched a takeover offer (OPA - {\em Offerta Pubblica d'Acquisto} in Italian) to buy Bulgari shares at
	12.25 euros. On that day, this share price jumped from below 8 euros to more than 12 euros.}
	\label{Tab:StatisticaPrice}
\begin{tabular}{|l|r|c|r|r|}
	\hline
  Assets  & mean $\times10^{-7}$ & variance $\times10^{-7}$ & skewness $\times10^{-2}$ & kurtosis \\
  \hline
A2A  &      $       29.15  $   &    $       5.24  $  &     $      9.36   $    &     $       5.22  $  \\
STS  &      $      -14.43  $   &    $       6.76  $  &     $     -7.11   $    &     $      11.50  $ \\
ATL  &      $       1.59   $   &    $       2.09  $  &     $      24.62  $    &     $      19.64  $ \\
AGL  &      $      -36.50  $   &    $       6.09  $  &     $      114.90 $    &     $      43.47  $ \\
AZM  &      $      -3.29   $   &    $       8.03  $  &     $     -21.90  $    &     $      14.14  $ \\
BP   &      $      -4.53   $   &    $       4.55  $  &     $     -1.69   $    &     $      10.69  $ \\
BMPS &      $       24.93  $   &    $       4.79  $  &     $     -21.71  $    &     $      24.34  $ \\
PMI  &      $       6.87   $   &    $       5.55  $  &     $     -23.73  $    &     $      41.72  $ \\
BUL (*)  &      $      -3.75   $   &    $       4.37  $  &     $     -295.68 $    &     $     154.69  $ \\
BZU  &      $       61.92  $   &    $       7.41  $  &     $     -99.04  $    &     $      35.92  $ \\
CPR  &      $       2.35   $   &    $       3.73  $  &     $      11.04  $    &     $       8.13  $ \\
DIA  &      $      -40.04  $   &    $       4.42  $  &     $     -49.99  $    &     $      29.17  $ \\
ENEL &      $       6.21   $   &    $       1.38  $  &     $      140.10 $    &     $      76.06  $ \\
EGPW &      $       38.81  $   &    $       3.64  $  &     $      3.43   $    &     $       7.31  $ \\
ENI  &      $       7.86   $   &    $       1.40  $  &     $      59.89  $    &     $      21.01  $ \\
EXO  &      $       11.98  $   &    $       4.82  $  &     $     -5.45   $    &     $       8.06  $ \\
F    &      $      -3.55   $   &    $       2.81  $  &     $     -45.05  $    &     $      21.76  $ \\
FI   &      $       14.33  $   &    $       3.68  $  &     $     -39.37  $    &     $      18.14  $ \\
FNC  &      $       0.50   $   &    $       3.29  $  &     $      28.01  $    &     $      13.01  $ \\
FSA  &      $       84.68  $   &    $       10.35 $  &     $     -163.51 $    &     $     180.64  $ \\
G    &      $       5.03   $   &    $       2.09  $  &     $     -100.65 $    &     $      44.97  $ \\
IPG  &      $       80.66  $   &    $       9.04  $  &     $     -45.81  $    &     $      22.68  $ \\
ISP  &      $       1.99   $   &    $       3.45  $  &     $     -62.87  $    &     $      43.12  $ \\
LTO  &      $       67.82  $   &    $       9.28  $  &     $     -171.44 $    &     $      62.62  $ \\
LUX  &      $       25.88  $   &    $       2.67  $  &     $      30.48  $    &     $      24.43  $ \\
MS   &      $       5.76   $   &    $       2.86  $  &     $     -22.98  $    &     $      19.38  $ \\
MB   &      $       17.29  $   &    $       4.18  $  &     $      1.66   $    &     $       9.67  $ \\
MED  &      $       20.25  $   &    $       7.64  $  &     $     -43.78  $    &     $      18.78  $ \\
PLT  &      $       9.76   $   &    $       5.30  $  &     $      49.56  $    &     $      14.43  $ \\
PC   &      $       47.93  $   &    $       5.41  $  &     $      3.44   $    &     $      10.75  $ \\
PRY  &      $       21.54  $   &    $       4.02  $  &     $      257.09 $    &     $      92.76  $ \\
SPM  &      $       5.72   $   &    $       1.50  $  &     $     -9.12   $    &     $      32.75  $ \\
SRG  &      $       12.09  $   &    $       2.41  $  &     $      79.03  $    &     $      54.87  $ \\
STM  &      $       15.69  $   &    $       2.56  $  &     $     -39.64  $    &     $      36.78  $ \\
TIT  &      $       8.33   $   &    $       3.20  $  &     $     -22.22  $    &     $       8.92  $ \\
TEN  &      $       0.34   $   &    $       2.61  $  &     $     -112.99 $    &     $     135.05  $ \\
TRN  &      $       26.67  $   &    $       2.42  $  &     $      3.54   $    &     $       6.03  $ \\
TOD  &      $       28.73  $   &    $       6.95  $  &     $      158.96 $    &     $      86.49  $ \\
UBI  &      $      -1.76   $   &    $       4.99  $  &     $     -67.53  $    &     $      25.23  $ \\
UCG  &      $       3.44   $   &    $       1.29  $  &     $     -12.56  $    &     $      57.51  $ \\
\hline
Index&      $1.10$      &    $0.03$   &      $2$    &     $8.54$    \\
  \hline
\end{tabular}
\end{table*}

\begin{table*}[tb]\footnotesize
	\centering
	\caption{Kolmogorov-Smirnov test, Jarque-Bera test and Lilliefors test to test the nomality of the empirical distributions corresponding to $\Delta t$ equal to 3s, 5s, 10s, 30s, 300s. The null hypothesis is always rejected. }
	\label{Fig7a}
\begin{tabular}{|l|r|r|r|}
	\hline
  $\Delta t$ & K-S test & J-B test & Lilliefors test\\
  \hline
\input{table_Fig7}
  \hline
\end{tabular}
\end{table*}

\begin{table*}[tb]\footnotesize
	\centering
	\caption{Kolmogorov-Smirnov test. In bold the rejected cases. The null hypothesis of empirical data coming from an identical distribution is rejected in the comparisons of $\Delta t = 3s$ and $\Delta t = 5s$, $\Delta t = 3s$ and $\Delta t = 10s$ and $\Delta t = 3s$ and $\Delta t = 30s$. }
	\label{Tab:Fig7c}
\begin{tabular}{|l|r|r|r|r|r|}
	\hline
  $\Delta t$ & 3s & 5s & 10s & 30s & 300s\\
  \hline
\input{table_kstest_Fig7c_norm}
  \hline
\end{tabular}
\end{table*}

\begin{table*}[tb]\footnotesize
	\centering
	\caption{Kolmogorov-Smirnov test, Jarque-Bera test and Lilliefors test for the normality of simulated data }
	\label{Fig14a}
\begin{tabular}{|l|r|r|r|}
	\hline
  $\Delta t$ & K-S test & J-B test & Lilliefors test\\
  \hline
\input{table_Fig14}
  \hline
\end{tabular}
\end{table*}
\begin{table*}[tb]\footnotesize
	\centering
	\caption{Kolmogorov-Smirnov test. The null hypothesis of simulated data coming from an identical distribution is always rejected.}
	\label{Tab:Fig14c}
\begin{tabular}{|l|r|r|r|r|r|}
	\hline
  $\Delta t$ & 3s & 5s & 10s & 30s & 300s\\
  \hline
\input{table_kstest_Fig14c_norm}
  \hline
\end{tabular}
\end{table*}

\begin{table}[htbp]
\caption{Parameter settings for the simulation of ACD data}
\begin{center}
\begin{tabular}{|l|r|r|l|r|l|}
\hline
 & \multicolumn{1}{l|}{$\omega$} & \multicolumn{1}{l|}{$\alpha_1$} & $\alpha_2$ & \multicolumn{1}{l|}{$\beta_1$} & $\beta_2$ \\ \hline
ACD(1,1) & 1 & 0.089 & -- & 0.85 & -- \\ \hline
ACD(1,2) & 1 & 0.1 & -- & 0.45 & \multicolumn{1}{r|}{0.4} \\ \hline
ACD(2,1) & 1 & 0.15 & \multicolumn{1}{r|}{0.15} & 0.65 & -- \\ \hline
ACD(2,2) & 1 & 0.1 & \multicolumn{1}{r|}{0.1} & 0.42 & \multicolumn{1}{r|}{0.35} \\ \hline
\end{tabular}
\end{center}
\label{acdParam}
\end{table}

\begin{table}[htbp]
\centering
\caption{Table of summary statistics of the MSE of the parameters $\mu$ and $\sigma^2$ of the compound Poisson type model. The analysis is based on 1000 samples generated from a simulation grid containing 30 subintervals.}
\begin{tabular}{|l|r|r|r|r|}
\hline
 & \multicolumn{1}{l|}{mean} & \multicolumn{1}{l|}{min} & \multicolumn{1}{l|}{max} & \multicolumn{1}{l|}{std} \\ \hline
$\mu$ & 0.0545 & 0.0026 & 0.1049 & 0.0212 \\ \hline
$\sigma^2$ & 0.1038 & 0.0049 & 0.1757 & 0.0439 \\ \hline
\end{tabular}
\label{mseMuSIgma}
\end{table}

\begin{table}[htbp]
\caption{Results of the MSE calculations for the ACD model}
\begin{center}
\begin{tabular}{|l|l||r|r|r|r|r|}
\hline
 &  & \multicolumn{1}{l|}{$\text{MSE}(\omega)$} & \multicolumn{1}{l|}{$\text{MSE}(\alpha_1)$} & \multicolumn{1}{l|}{$\text{MSE}(\alpha_2)$} & \multicolumn{1}{l|}{$\text{MSE}(\beta_1)$} & \multicolumn{1}{l|}{$\text{MSE}(\beta_2)$} \\ \hline\hline
ACD(1,1) & T=250 & 3.7508 & 0.0023 & -- & 0.0231 & -- \\ 
 & T=500 & 1.8887 & 0.0010 & -- & 0.0108 & -- \\ 
 & T=1000 & 0.3591 & 0.0005 & -- & 0.0025 & -- \\ 
 & T=2000 & 0.1245 & 0.0002 & -- & 0.0010 & -- \\ \hline
ACD(1,2) & T=250 & 14.5255 & 0.0036 & -- & 0.4748 & 0.4282 \\ 
 & T=500 & 3.7468 & 0.0019 & -- & 0.3039 & 0.2681 \\ 
 & T=1000 & 0.6259 & 0.0010 & -- & 0.1869 & 0.1606 \\ 
 & T=2000 & 0.1905 & 0.0005 & -- & 0.0809 & 0.0681 \\ \hline
ACD(2,1) & T=250 & 0.8491 & 0.0063 & 0.0108 & 0.0130 & -- \\ 
 & T=500 & 0.2664 & 0.0032 & 0.0050 & 0.0053 & -- \\ 
 & T=1000 & 0.0916 & 0.0014 & 0.0026 & 0.0023 & -- \\ 
 & T=2000 & 0.0418 & 0.0007 & 0.0012 & 0.0011 & -- \\ \hline
ACD(2,2) & T=250 & 6.4135 & 0.0067 & 0.0102 & 0.3165 & 0.2445 \\ 
 & T=500 & 1.1077 & 0.0032 & 0.0061 & 0.2722 & 0.2031 \\ 
 & T=1000 & 0.3730 & 0.0014 & 0.0041 & 0.2086 & 0.1526 \\ 
 & T=2000 & 0.1512 & 0.0006 & 0.0026 & 0.1612 & 0.1181 \\ \hline
\end{tabular}
\end{center}
\label{mseACD}
\end{table}

\begin{table}[htbp]
\center
\caption{Model selection results based on ACD(1,1) data samples: Given 1000 samples of size $T\in \{250, 500, 1000, 2000\}$ each column gives the percentage of cases in which the different IC selected the models ACD(1,1), ACD(1,2), ACD(2,1) and ACD(2,2) respectively. The bold numbers give the largest percentage per row.}
\begin{tabular}{|l|l||r|r|r|r|}
\hline
 &  & \multicolumn{1}{l|}{ACD(1,1)} & \multicolumn{1}{l|}{ACD(1,2)} & \multicolumn{1}{l|}{ACD(2,1)} & \multicolumn{1}{l|}{ACD(2,2)} \\ \hline\hline
T=250 & AIC & \textbf{58.7} & 23.6 & 9.9 & 7.8 \\ 
 & BIC & \textbf{90.2} & 7 & 2.1 & 0.7 \\ 
 & HQ & \textbf{77.9} & 14.6 & 4.8 & 2.7 \\ \hline
T=500 & AIC & \textbf{62.9} & 20.4 & 10.9 & 5.8 \\ 
 & BIC & \textbf{93.6} & 4.7 & 1.6 & 0.1 \\ 
 & HQ & \textbf{82.6} & 11.5 & 4.9 & 1 \\ \hline
T=1000 & AIC & \textbf{67.5} & 16.4 & 11 & 5.1 \\ 
 & BIC & \textbf{97.4} & 1.8 & 0.8 & 0 \\ 
 & HQ & \textbf{87.2} & 7.5 & 4.8 & 0.5 \\ \hline
T=2000 & AIC & \textbf{71.3} & 13.1 & 9.7 & 5.9 \\ 
 & BIC & \textbf{97.7} & 1.6 & 0.6 & 0.1 \\ 
 & HQ & \textbf{91.5} & 4.4 & 3 & 1.1 \\ \hline
\end{tabular}
\label{ACD11}
\end{table}

\begin{table}[htbp]
\center
\caption{Model selection results based on ACD(1,2) data samples: Given 1000 samples of size $T\in \{250, 500, 1000, 2000\}$ each column gives the percentage of cases in which the different IC selected the models ACD(1,1), ACD(1,2), ACD(2,1) and ACD(2,2) respectively. The bold numbers give the largest percentage per row.}
\begin{tabular}{|l|l||r|r|r|r|}
\hline
 &  & \multicolumn{1}{l|}{ACD(1,1)} & \multicolumn{1}{l|}{ACD(1,2)} & \multicolumn{1}{l|}{ACD(2,1)} & \multicolumn{1}{l|}{ACD(2,2)} \\ \hline\hline
T=250 & AIC & \textbf{58.6} & 24.7 & 9.6 & 7.1 \\ 
 & BIC & \textbf{91.5} & 6.5 & 1.3 & 0.7 \\ 
 & HQ & \textbf{78.6} & 14.8 & 3.7 & 2.9 \\ \hline
T=500 & AIC & \textbf{60.6} & 25.1 & 10.3 & 4 \\ 
 & BIC & \textbf{94.7} & 4.3 & 0.7 & 0.3 \\ 
 & HQ & \textbf{81.2} & 13.5 & 4.5 & 0.8 \\ \hline
T=1000 & AIC & \textbf{52.7} & 27.8 & 15.2 & 4.3 \\ 
 & BIC & \textbf{92.6} & 5.1 & 2.3 & 0 \\ 
 & HQ & \textbf{76} & 14.7 & 8.8 & 0.5 \\ \hline
T=2000 & AIC & \textbf{41.5} & 35.6 & 18 & 4.9 \\ 
 & BIC & \textbf{88.4} & 6.7 & 4.9 & 0 \\ 
 & HQ & \textbf{67.6} & 20.4 & 11.6 & 0.4 \\ \hline
\end{tabular}
\label{ACD12}
\end{table}

\begin{table}[htbp]
\center
\caption{Model selection results based on ACD(2,1) data samples: Given 1000 samples of size $T\in \{250, 500, 1000, 2000\}$ each column gives the percentage of cases in which the different IC selected the models ACD(1,1), ACD(1,2), ACD(2,1) and ACD(2,2) respectively. The bold numbers give the largest percentage per row.}
\begin{tabular}{|l|l||r|r|r|r|}
\hline
 &  & \multicolumn{1}{l|}{ACD(1,1)} & \multicolumn{1}{l|}{ACD(1,2)} & \multicolumn{1}{l|}{ACD(2,1)} & \multicolumn{1}{l|}{ACD(2,2)} \\ \hline\hline
T=250 & AIC & \textbf{36.2} & 20.9 & 31.8 & 11.1 \\ 
 & BIC & \textbf{73.7} & 8.9 & 16.8 & 0.6 \\ 
 & HQ & \textbf{52.4} & 16.3 & 28.1 & 3.2 \\ \hline
T=500 & AIC & 19.1 & 20.7 & \textbf{50} & 10.2 \\ 
 & BIC & \textbf{59.9} & 10.5 & 29 & 0.6 \\ 
 & HQ & 36.5 & 16.4 & \textbf{43.8} & 3.3 \\ \hline
T=1000 & AIC & 7.4 & 16.7 & \textbf{64.8} & 11.1 \\ 
 & BIC & 35.6 & 11.9 & \textbf{52.1} & 0.4 \\ 
 & HQ & 17.1 & 15.7 & \textbf{63.7} & 3.5 \\ \hline
T=2000 & AIC & 1.2 & 12.7 & \textbf{74.2} & 11.9 \\ 
 & BIC & 6.8 & 12.9 & \textbf{80.1} & 0.2 \\ 
 & HQ & 2.2 & 14.2 & \textbf{81.6} & 2 \\ \hline
\end{tabular}
\label{ACD21}
\end{table}

\begin{table}[htbp]
\center
\caption{Model selection results based on ACD(2,2) data samples: Given 1000 samples of size $T\in \{250, 500, 1000, 2000\}$ each column gives the percentage of cases in which the different IC selected the models ACD(1,1), ACD(1,2), ACD(2,1) and ACD(2,2) respectively. The bold numbers give the largest percentage per row.}
\begin{tabular}{|l|l||r|r|r|r|}
\hline
 &  & \multicolumn{1}{l|}{ACD(1,1)} & \multicolumn{1}{l|}{ACD(1,2)} & \multicolumn{1}{l|}{ACD(2,1)} & \multicolumn{1}{l|}{ACD(2,2)} \\ \hline\hline
T=250 & AIC & \textbf{56.7} & 15.8 & 18.8 & 8.7 \\ 
 & BIC & \textbf{89.7} & 5.3 & 4.5 & 0.5 \\ 
 & HQ & \textbf{74} & 11.5 & 11.7 & 2.8 \\ \hline
T=500 & AIC & \textbf{57.2} & 13.6 & 19.1 & 10.1 \\ 
 & BIC & \textbf{92.1} & 2.9 & 4.6 & 0.4 \\ 
 & HQ & \textbf{78.4} & 8 & 11.4 & 2.2 \\ \hline
T=1000 & AIC & \textbf{48.4} & 13.1 & 23.4 & 15.1 \\ 
 & BIC & \textbf{91.5} & 2.7 & 5.7 & 0.1 \\ 
 & HQ & \textbf{74} & 6.9 & 16.1 & 3 \\ \hline
T=2000 & AIC & 34.2 & 9.7 & \textbf{37.2} & 18.9 \\ 
 & BIC & \textbf{86.1} & 1.8 & 11.5 & 0.6 \\ 
 & HQ & \textbf{59.7} & 6.8 & 26.5 & 7 \\ \hline
\end{tabular}
\label{ACD22}
\end{table}

\nolinenumbers

%
%
%
%
%
%
%

\end{document}

%% file: table1newtest.tex
\begin{table*}[tb]\footnotesize
	\centering
	\caption{Descriptive statistics for the waiting times $\tau^h$}
	\label{Tab:StatisticaTempiAttesa}
\begin{tabular}{|l|r|r|r|r|r|r|}
	\hline
  Asset & mean & std & $\alpha$ & $\beta$ &   AD & Lillie\\
  \hline
A2A   &        32.49   &       39.04  & 0.053 & 0.865& 106  & 0.068    \\
STS   &        34.07   &       43.68  & 0.061 & 0.818& 122  & 0.083    \\
ATL   &        24.42   &       32.48  & 0.088 & 0.792& 263  & 0.099    \\
AGL   &        33.20   &       41.87  & 0.059 & 0.830& 145  & 0.082    \\
AZM   &        34.67   &       42.35  & 0.052 & 0.853& 116  & 0.074    \\
BP    &         9.54   &       12.80  & 0.189 & 0.786& 1158 & 0.134    \\
BMPS  &        17.21   &       23.96  & 0.130 & 0.761& 401  & 0.107    \\
PMI   &        19.95   &       27.26  & 0.111 & 0.773& 293  & 0.099    \\
BUL   &        24.87   &       37.02  & 0.116 & 0.717& 326  & 0.123    \\
BZU   &        22.62   &       33.71  & 0.125 & 0.716& 435  & 0.125    \\
CPR   &        33.77   &       42.42  & 0.058 & 0.833& 174  & 0.092    \\
DIA   &        30.21   &       39.91  & 0.073 & 0.797& 155  & 0.091    \\
ENEL  &         9.19   &       11.60  & 0.173 & 0.829& 987  & 0.123    \\
EGPW  &        21.16   &       29.31  & 0.110 & 0.764& 239  & 0.094    \\
ENI   &         8.71   &       12.21  & 0.221 & 0.756& 1541 & 0.148    \\
EXO   &        22.72   &       31.16  & 0.101 & 0.771& 228  & 0.094    \\
F     &         7.94   &       11.29  & 0.243 & 0.747& 1936 & 0.158    \\
FI    &        12.80   &       18.77  & 0.182 & 0.726& 833  & 0.132    \\
FNC   &        20.86   &       26.98  & 0.093 & 0.812& 234  & 0.089    \\
FSA   &        23.70   &       35.15  & 0.120 & 0.719& 309  & 0.118    \\
G     &        11.10   &       14.79  & 0.165 & 0.792& 759  & 0.119    \\
IPG   &        32.26   &       41.41  & 0.064 & 0.818& 157  & 0.085    \\
ISP   &         7.96   &       11.30  & 0.242 & 0.748& 1930 & 0.158    \\
LTO   &        33.22   &       42.54  & 0.062 & 0.819& 117  & 0.082    \\
LUX   &        23.28   &       31.52  & 0.096 & 0.780& 231  & 0.096    \\
MS    &        20.12   &       27.93  & 0.114 & 0.763& 350  & 0.107    \\
MB    &        17.40   &       24.03  & 0.126 & 0.767& 403  & 0.108    \\
MED   &        31.66   &       39.57  & 0.060 & 0.837& 126  & 0.077    \\
PLT   &        20.49   &       29.01  & 0.119 & 0.749& 322  & 0.104    \\
PC    &        22.78   &       30.45  & 0.094 & 0.789& 221  & 0.092    \\
PRY   &        19.48   &       27.87  & 0.126 & 0.743& 390  & 0.113    \\
SPM   &        11.53   &       17.88  & 0.219 & 0.691& 1185 & 0.150    \\
SRG   &        24.77   &       32.77  & 0.086 & 0.796& 208  & 0.091    \\
STM   &        12.22   &       17.26  & 0.174 & 0.751& 750  & 0.124    \\
TIT   &        13.27   &       20.52  & 0.198 & 0.692& 972  & 0.146    \\
TEN   &        17.49   &       24.98  & 0.137 & 0.743& 395  & 0.110    \\
TRN   &        28.12   &       35.52  & 0.068 & 0.829& 148  & 0.080    \\
TOD   &        31.31   &       40.71  & 0.068 & 0.808& 114  & 0.081    \\
UBI   &        20.58   &       27.30  & 0.100 & 0.794& 272  & 0.096    \\
UCG   &         3.85   &        4.94  & 0.364 & 0.817& 8640 & 0.223    \\
\hline
Index &         1.66   &        1.26   &    --     &  --     & Inf  & 0.365\\
  \hline
\end{tabular}
\end{table*}

%% file: table_Fig7.tex
 3s & 0.499 & 3108277492760.052 & 0.135 \\
 5s &0.499 & 1087100884817.007 & 0.142  \\
 10s & 0.499 & 117920812948.739 & 0.141  \\
 30s & 0.498 & 3409200688.215 & 0.128   \\
 300s &0.497 & 7949646.601 & 0.136   \\

%% file: table_kstest_Fig7c_norm.tex
 3s & -- & \textbf{0.010} & \textbf{0.014} &\textbf{0.014} & 0.023 \\
 5s &\textbf{0.010} & -- & 0.008 &0.010 & 0.022 \\
 10s & \textbf{0.014} & 0.008 & --&0.008 & 0.017  \\
 30s & \textbf{0.014} & 0.010 & 0.008 &-- & 0.018  \\
 300s &0.023 & 0.022 & 0.017 &0.018 & --  \\ 

%% file: table_Fig14.tex
 3s & 0.499 & 12727206295498.855 & 0.209 \\
 5s &0.499 & 2484970052007.152 & 0.200  \\
 10s & 0.499 & 279530930024.888 & 0.189  \\
 30s & 0.498 & 9268127864.106 & 0.173   \\
 300s &0.490 & 8190272.873 & 0.157   \\

%% file: table_kstest_Fig14c_norm.tex
 3s & -- & 0.019 & 0.031 &0.036 & 0.035 \\
 5s &0.019 & -- & 0.012 &0.018 & 0.018 \\
 10s & 0.031 & 0.012 &--&0.007 & 0.016  \\
 30s & 0.036 & 0.018 & 0.007 &-- & 0.019  \\
 300s &0.035 & 0.018 & 0.016 &0.019 & --  \\ 